\documentclass[aps,pra,reprint,superscriptaddress, longbibliography]{revtex4-2}
\usepackage{lipsum, babel}

\usepackage{siunitx}
\usepackage{amsmath,bm}
\usepackage{amsbsy}
\usepackage{amssymb}
\usepackage{mathptmx, textcomp}
\usepackage{color}
\usepackage{braket}
\usepackage{graphicx}
\usepackage{textcomp}
\usepackage{notes2bib}
\usepackage{textcomp}
\usepackage{mwe}
\usepackage{ulem}

\definecolor{bl}{rgb}{0, .1, .6}
\usepackage[colorlinks=true, citecolor = bl, linkcolor = bl, urlcolor=bl, pdfborder={0 0 0}]{hyperref}

\begin{document}
\title{Storage and release  of subradiant excitations in a dense atomic cloud}

\author{Giovanni Ferioli}
\author{Antoine Glicenstein}
\affiliation{Universit\'e Paris-Saclay, Institut d'Optique Graduate School, CNRS, 
Laboratoire Charles Fabry, 91127, Palaiseau, France}
\author{Loic Henriet}
\affiliation{Pasqal, 2 avenue Augustin Fresnel, 91120 Palaiseau, France}
\author{Igor Ferrier-Barbut}
\email{igor.ferrier-barbut@institutoptique.fr}
\author{Antoine Browaeys}
\affiliation{Universit\'e Paris-Saclay, Institut d'Optique Graduate School, CNRS, 
Laboratoire Charles Fabry, 91127, Palaiseau, France}

\begin{abstract}
We report the observation of subradiance in dense ensembles of cold $^{87}$Rb atoms operating near Dicke's regime of a large number of atoms in a volume with dimensions smaller than the transition wavelength.  We validate that the atom number is the only cooperativity parameter governing subradiance. We probe the dynamics in the many-body regime and support the picture that multiply-excited subradiant states are built as a superposition of singly-excited states that decay independently. Moreover, we implement an experimental procedure to release the 
excitation stored in the long-lived modes in a pulse of light. 
This technique is a first step towards the realization of tailored light 
storing based on subradiance.  
\end{abstract}

\maketitle
The interaction between a single two-level atom and radiation is well understood:
the atomic response is described by a resonance frequency and a decay rate. 
When considering more than one emitter in a volume with dimensions smaller than the transition wavelength, this response may be altered, 
as was proposed in the pioneering work of Dicke \citep{dicke1954, gross1982}. 
Indeed, light-induced interactions modify dramatically the behavior of the ensemble and its response becomes collective. 
In particular, the decay rate of excitations hosted in the ensemble can be starkly modified. 
Superradiance, i.e.~a decay of the excitation at a rate faster than the single-atom one, 
has been verified experimentally in atomic systems from ions  to dilute clouds of atoms 
\cite{gross1982,devoe1996observation,araujo2016superradiance, roof2016observation,solano2017super}. 
The study of its counterpart, namely subradiance with a decay rate smaller than the atomic one, 
has been restricted to a handful of works \cite{pavolini1985experimental}: 
direct observations were reported in a pair of ions at variable distance 
\cite{devoe1996observation} and in molecular systems
\cite{Hettich2002,takasu2012controlled,mcguyer2015precise}. 
Recently, it was also observed in a cold, dilute 
atomic cloud \cite{guerin2016subradiance, cipris2020subradiance, das2020subradiance} 
and  as a line-narrowing in an ordered 2D layer of atoms \cite{rui2020subradiant}.\par

Engineering subradiant states has drawn an increasing attention since 
it might pave the way to several applications. For instance, the possibility to 
store an excitation in subradiant modes and to address it in real time while the 
excitation is stored has inspired proposals to use it as a storage medium 
\cite{plankensteiner2015selective,facchinetti2016storing,PhysRevA.94.013803,
asenjo2017exponential,needham2019subradiance}. 
Secondly, the narrowing of the line associated to subradiant modes and their subsequent 
enhanced sensitivity to external fields could be a promising 
application for metrology \cite{PhysRevLett.111.123601,plankensteiner2015selective,facchinetti2018interaction}. 
Recent proposals have also suggested to use subradiance as a tool for quantum information processing and 
quantum optics \cite{wild2018quantum,guimond2019subradiant}.\par

All these proposals have been formulated in ordered systems 
with small inter-particle distances, $\bar r\lesssim\lambda$ where $\lambda$ 
is the wavelength of the atomic transition. 
Motivated by this, we take a first step in this direction by 
exploring the regime $\bar r<\lambda$, but in the disordered case 
using dense clouds of $^{87}$Rb atoms, characterized by a peak density $\rho_0$ satisfying 
$\rho_0\lambda^3\gg 1$ ($\bar r=\rho_0^{-1/3}$). 
Furthermore, the ensembles we produce have a prolate shape with 
typical radial size of $\sim0.5\,\lambda$ and an axial one of $\sim5\,\lambda$. 
We thus closely approach Dicke's regime where many emitters are trapped in 
a volume comparable to the wavelength of their transition.  This regime introduces several important differences with respect to the case 
of a dilute extended cloud studied in
refs.~\cite{guerin2016subradiance, cipris2020subradiance, das2020subradiance}. 
Firstly, here the parameter governing the collective properties is 
no longer the optical depth, but rather the number of atoms in the cloud. 
Indeed the cooperativity $C$ is the ratio between the atom number $N$ and the number of modes 
efficiently coupled to the system $M$, i.e., $C=N/M$ \cite{guerin2017light}. 
In a cloud with a volume much larger than $\lambda^3$, this parameter 
is  the optical depth on resonance, which was experimentally shown to govern   
collective effects \cite{guerin2016subradiance,araujo2016superradiance}. 
Approaching Dicke's regime, as we do here, the ensemble becomes  efficiently 
coupled only to a single mode and thus the cooperativity
should be the atom number $N$ \cite{akkermans2008photon}.   
Secondly, since in our clouds $k\bar r\sim1$ ($k=2\pi/\lambda$), 
all the terms of the dipole-dipole interaction \cite{jackson2007classical} play a role, 
as opposed to the dilute regime where only a radiative $1/r$ term is considered \cite{guerin2017light}. 

Here we observe subradiance in the time domain in a cloud operating near Dicke's regime. First we validate the characteristic dependence on the atom number. Second we explore the storage of light in long-lived multiply-excited states. Varying the intensity of the excitation laser, we characterize these subradiant states containing few to many excitations. Our finding supports the idea that multiply-excited subradiant states are built as a superposition of singly-excited states in random ensembles, similarly to what was recently predicted in ordered 1D systems \cite{asenjo2017exponential,Albrecht2019Subradiant,henriet2019critical,zhang2020theory}. Finally, we demonstrate dynamical control of subradiance while excitations are stored by releasing it on demand via the application of a laser. 
This real-time control of the coupling of an ensemble to the 
electromagnetic modes while it hosts an excitation offers new possibilities for light storage.\par
\begin{figure*}
\centering
\hspace*{.05\linewidth}
\includegraphics[width=0.3\linewidth]{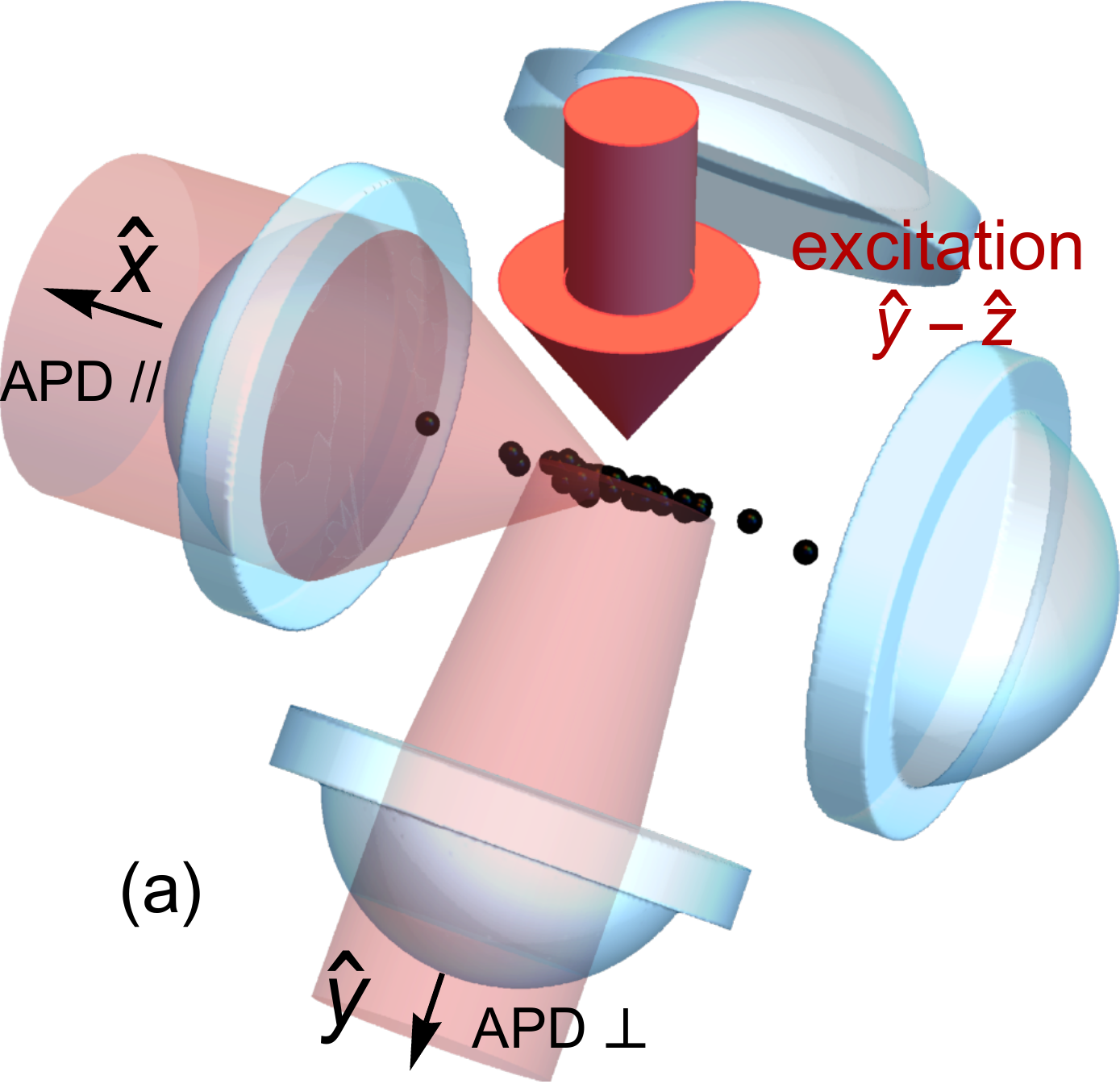}\hfil
\hspace*{.05\linewidth}
\includegraphics[width=0.4\linewidth]{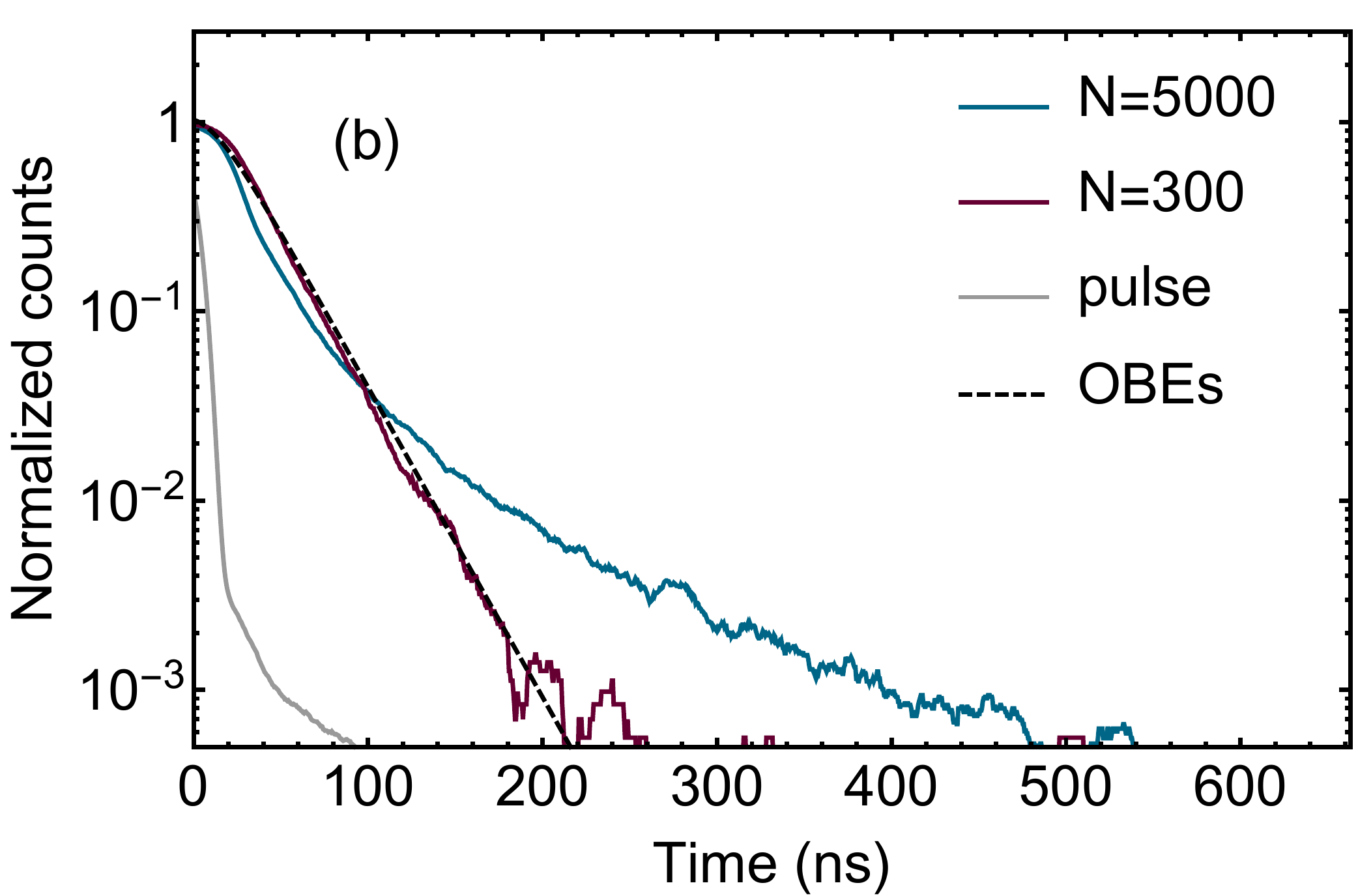}\par\medskip
\includegraphics[width=0.4\linewidth]{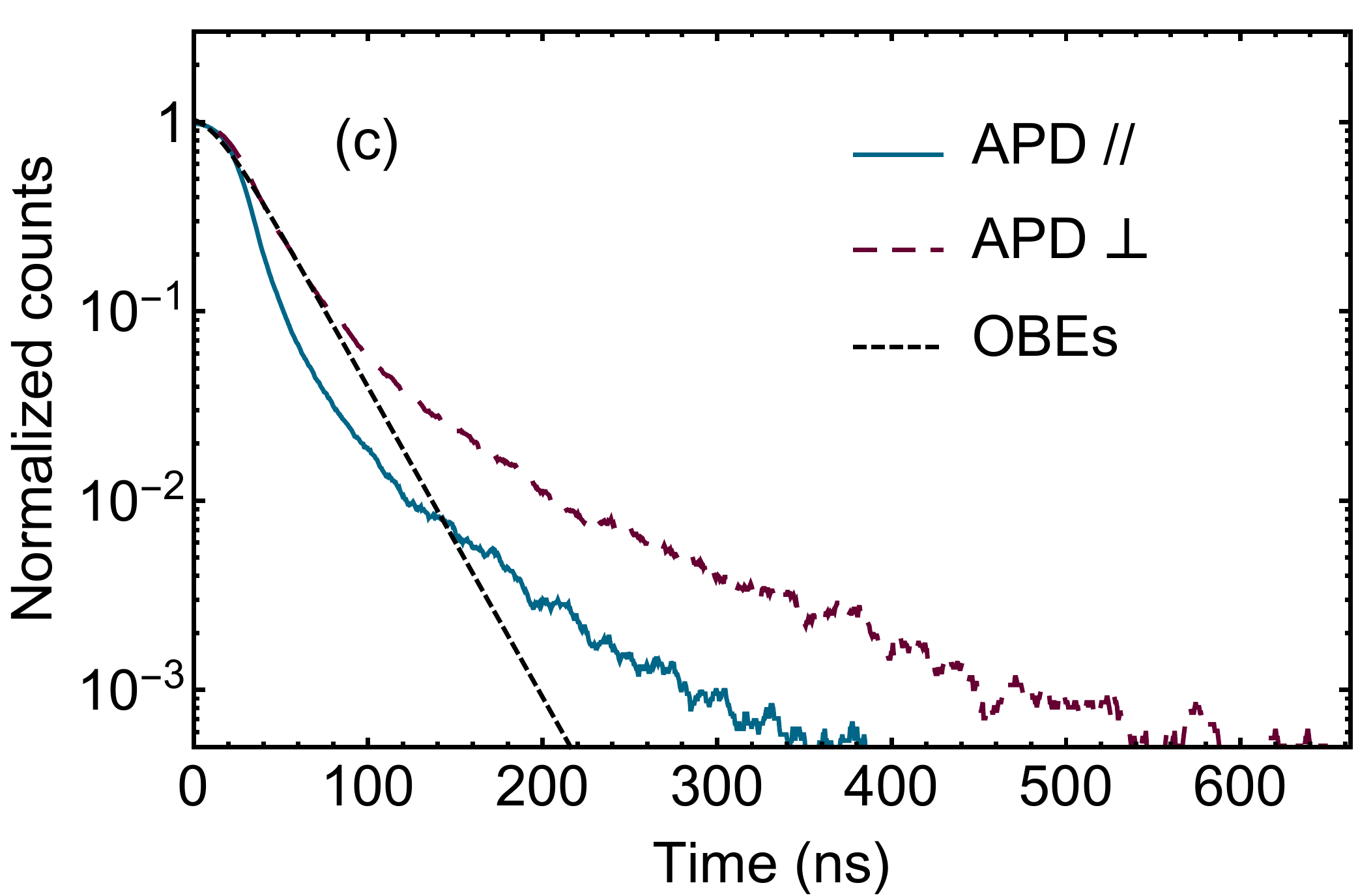}\hfil
\includegraphics[width=0.4\linewidth]{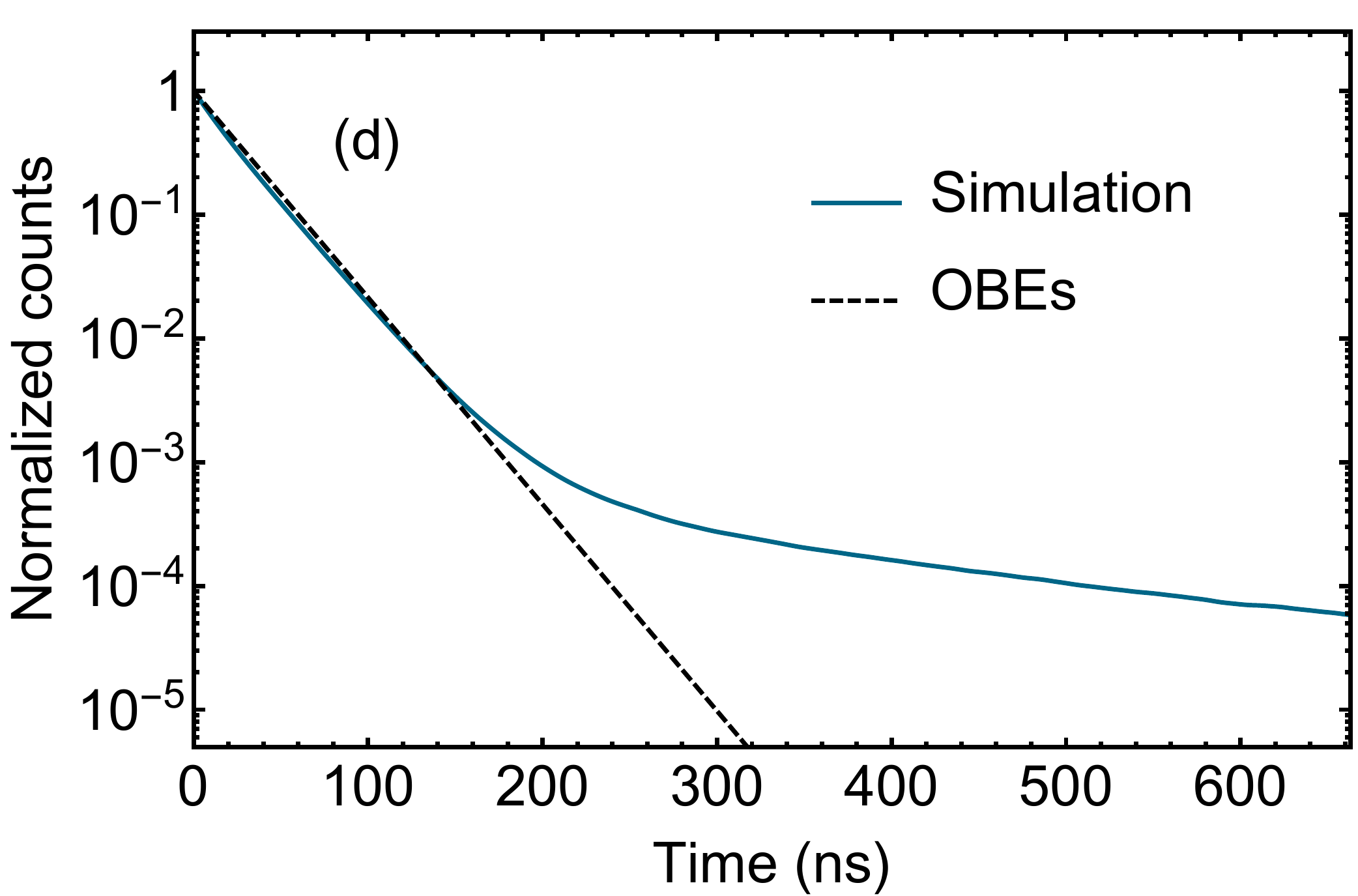}
\caption{(a) Schematics of the experimental set up.  Four high numerical aperture (0.5) lenses collect the fluorescence emitted by the atomic cloud along two axes, which is then fiber-coupled to avalanche photo-diodes (APD). 
The excitation beam is aligned in the vertical direction. The trap beam (not shown) propagates along $x$, 
which is also the first collection axis. 
The second collection axis is at $45^\circ$ with respect to the excitation direction. 
(b) Number of photons collected in bins of 0.5~ns as a function of time after switching off the excitation light (saturation parameter $s\simeq 27$)
for a cloud containing $\simeq5000$ (blue solid line) and $\simeq300$ atoms (purple solid line), data represented with a moving average.  This data is obtained by repetition of 20 pulses on 10000 clouds. 
Gray line: excitation pulse temporal shape.
Dashed line solution of the optical Bloch equations solved for our pulse shape. 
All curves have been normalized to their steady-state value during the excitation. 
(c) Time traces collected using the APD 
aligned along the $y$ direction (APD$\perp$, long-dashed purple), and along $x$ (APD\,// solid blue), the dashed black line shows the solution of OBEs. 
(d) Numerical simulations of the experiment using a non-linear coupled-dipole 
model \cite{do2020collective, Glicenstein2020} in an ensemble of 200 atoms 
with $\rho_0/k^3=0.3$. 
Black dashed line: solution of optical Bloch equations for a single atom. }
\label{fig1}
\end{figure*}

\section{Experimental setup}\label{sec:setup}

A detailed description of the experimental setup can be found elsewhere \cite{Setup, Glicenstein2020}. 
Briefly, as sketched in Fig.\,\ref{fig1}(a), it is composed of four aspherical lenses 
with large numerical aperture (0.5) in a maltese-cross configuration \cite{Bruno2019}. 
We use the $x$ high-resolution optical axis  to create 
an optical tweezer at a wavelength $\lambda_{\rm trap}=940$\,nm,
with a tunable waist (range 1.8-2.5 $\si{\micro\meter}$). 
Exploiting a gray-molasses loading on the D1 line, we trap $\simeq5000$ \textsuperscript{87}Rb atoms in the largest tweezer at a 
temperature of about $\SI{650}{\micro \kelvin}$ in a $\SI{4.2}{\milli \kelvin}$ trap. 
In this configuration, 
the trapping frequencies are $\omega_r\simeq 2 \pi \times \SI{81}{\kilo \hertz}$ 
and $\omega_z\simeq 2 \pi \times \SI{7}{\kilo \hertz}$, where $\omega_r$ and $\omega_z$ 
represent the radial and the axial directions. 
The central density of the cloud in these conditions is 
$\rho_0/k^3= \text{0.3} \pm \text{0.1}$ ($\bar r\simeq 0.2\,\lambda$), where $k=2\pi/\lambda$. 
The trap can then be compressed either by increasing the power 
of the trapping beam or reducing its waist \cite{Setup}. \par

We use the $F=2\to F'=3$ transition on the D2 line with wavelength 
$\lambda\simeq\SI{780}{\nano\meter}$, linewidth $\Gamma_0\simeq2\pi\times\SI{6}{\mega\hertz}$
and saturation intensity $I_{\rm sat}=1.6$\,mW/cm$^2$. 
In order to excite the cloud, we switch off the trap and shine a $\SI{150}{\nano \second}$-long pulse 
of resonant light along the $y-z$ direction of Fig.\,\ref{fig1}(a). This duration is long enough to reach the steady state during the excitation. After $\SI{1}{\micro \second}$, 
the atoms are re-captured in the tweezer. This time is short enough  ($<1/\omega_r$) that the density remains constant during this trap-free time. 
We repeat the same sequence up to 20 times on the same cloud (depending on the trap geometry), 
checking that the atom number is reduced by less than 10$\%$ at the end of the pulse sequence. We then repeat this sequence on 3000 to 10000 clouds at a $\SI{2}{\hertz}$ rate to obtain one trace (photon count per $\SI{0.5}{\nano\second}$ bin versus time). For the largest atom number, $N\simeq 5000$, we collect in the steady state typically $0.01$ photon per pulse in a $\SI{0.5}{\nano\second}$ time bin.
The excitation pulse is controlled by means of two acousto-optical modulators 
in series ensuring a high extinction ratio. 
The characteristic rise/fall time is $\SI{10}{\nano \second}$, the temporal shape of the pulse  is shown in Fig.\,\ref{fig1}(b) (gray line). 
The $\simeq$1 mm excitation beam waist is much larger than the cloud size 
so that all the atoms experience the same Rabi frequency. 
We can vary the saturation parameters $s = I/I_{\rm sat}$ up to $s\simeq 250$
with $I_{\rm sat}$ the saturation intensity (we have calibrated this intensity 
independently using dilute clouds). 
Exploiting the two high-resolution optical axes, we collect the fluorescence 
light into two fiber-coupled avalanche photodiodes (APDs) operating 
in single photon counting mode, one aligned along the axial direction of the cloud 
(APD\,//), the other perpendicularly to it (APD$\perp$), 
as sketched in Fig.\,\ref{fig1}(a).\par

\section{Subradiance near Dicke's regime}\label{sec:charac}

In Fig.\,\ref{fig1}(b) we report two time-resolved fluorescence traces recorded along $x$,
obtained for two clouds with respectively $\simeq$\,300 atoms 
($\rho_0/k^3\simeq0.02$, purple line) and $\simeq$\,5000 atoms ($\rho_0/k^3\simeq 0.3$, blue line) following the switch-off of the excitation laser for $s\simeq27$. 
We also show the solution of the optical Bloch equations (OBEs) for a single atom, 
solved for the measured pulse shape. In the low atom number case, the decay of 
the excitation is very well described by the OBEs, indicating that the atoms 
act as independent atoms. On the contrary for large atom number, 
the fluorescence decays non-exponentially: we observe first a decay at a rate larger 
than the single atom decay (superradiance), followed by a slower one (subradiance). 
Moreover, as shown in Fig.\,\ref{fig1}(c), 
we observe that superradiance occurs mainly in the axial direction of the cloud 
while the emission of subradiant excitation is observed in both directions. 
The finite switch-off time of the driving pulse limits the 
superradiant decay that can be observed. For this reason, in the rest of the paper 
we will focus our attention on the subradiant tail, leaving a detailed study of  
superradiance for future works. 

All the measurements reported here have 
been performed with resonant and linearly polarized light, in the absence of 
Zeeman optical pumping, thus exciting a multi-level system. 
We did not observe that the polarization direction impacts the 
observed subradiance. We have further observed that subradiance is essentially 
unchanged within our dynamical range in the presence of a magnetic field and a circularly polarized pulse 
with prior optical pumping (see appendix\,\ref{app:levStruc}). This suggests that the internal structure does 
not play a major role for subradiance in our regime.
Finally, we have checked that the subradiance is unchanged 
when we vary the detuning of the excitation laser around the atomic resonance 
(appendix\,\ref{app:det}). This indicates that radiation trapping of light in the cloud seen as a random
walk of the photons before escaping can not explain the observed slow decay \cite{labeyrie2003slow}.
For our small dense cloud, and contrarily to the case of dilute, optically 
thick clouds \cite{guerin2016subradiance}, this is expected, as  the photon mean-free path
$l_{\rm sc}=1/(\rho\sigma_{\rm sc})$ (with $\rho$ the atomic density and 
$\sigma_{\rm sc}=3\lambda^2/2\pi$ the resonant  cross-section) 
is smaller than the mean inter-particle distance. 

To further support our observations of subradiance, we perform numerical simulations in a simplified setting of two-level atoms. 
For the large number of atoms involved in the experiment, the ab-initio simulation  
by use of a master equation is beyond reach and we therefore resort to approximations. 
We use a non-linear coupled-dipoles model \cite{do2020collective,Glicenstein2020} 
consisting of a coupled system of OBEs, given in appendix\,\ref{app:NLCD}. It formally amounts to a mean-field approximation, 
assuming that the density matrix of the system can be factorized 
\cite{do2020collective,Kramer2015generalized,Parmee2018phases}. 
It allows one to take into account saturation effects of individual atoms. 
We numerically solved the equations for $N = 200$ atoms 
at a density $\rho_0/k^3=0.3$ [Fig.\,\ref{fig1}(d)]. 
The results do not feature superradiance, but do yield subradiance. 
The origin of the superradiance observed in the experiment 
and not present in the mean-field simulation is left for future investigations. 
The prediction of subradiance in our simulations 
could suggest that the mean-field model is enough to account for our observations. 
However as we will show below, it fails to reproduce our results 
in the strongly saturated regime, even qualitatively. 
In the remaining of this section, we characterize the observed 
subradiance as a function of atom number.\par
\begin{figure}
\includegraphics[width=\linewidth]{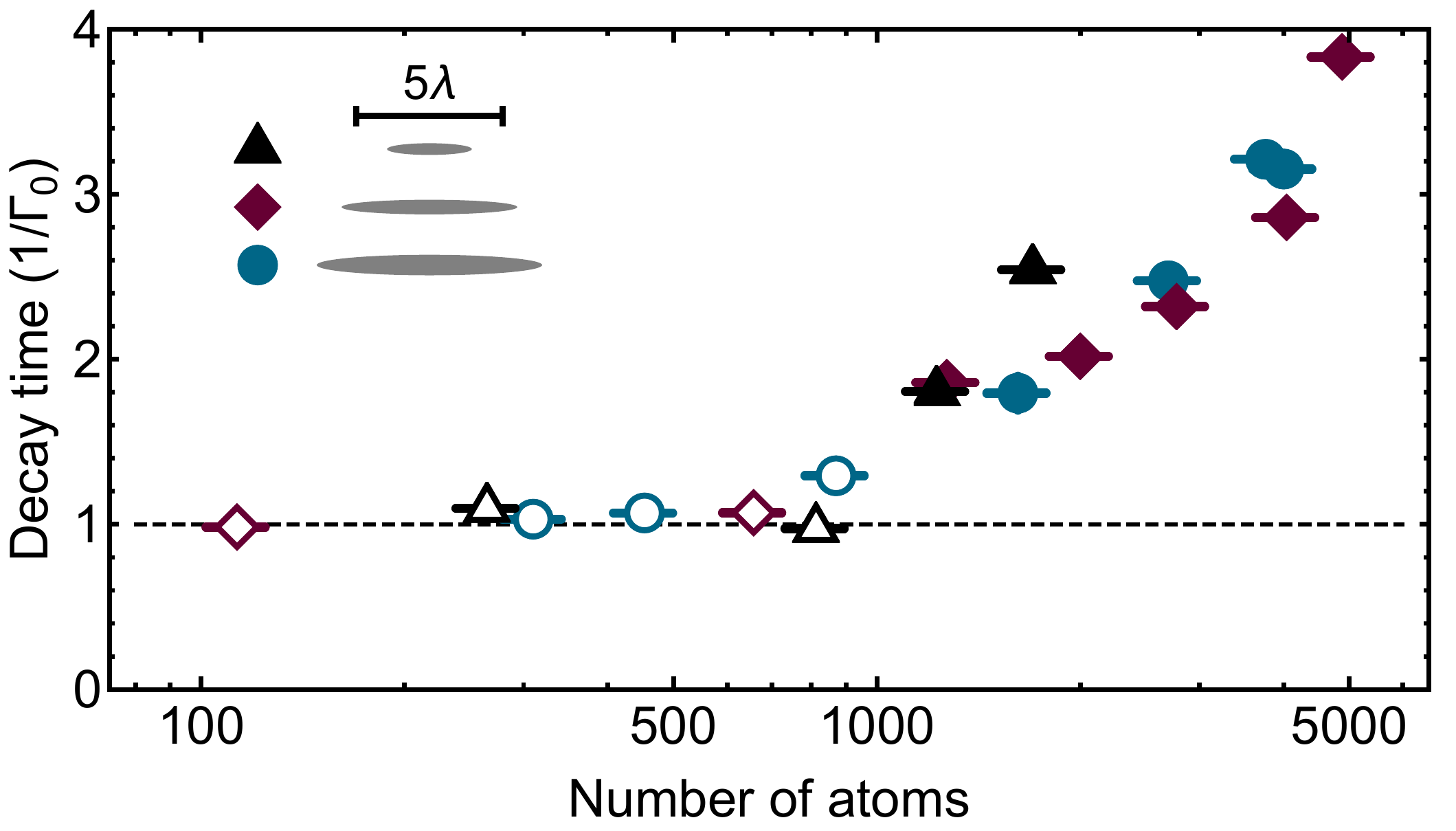} 
\includegraphics[width=\linewidth]{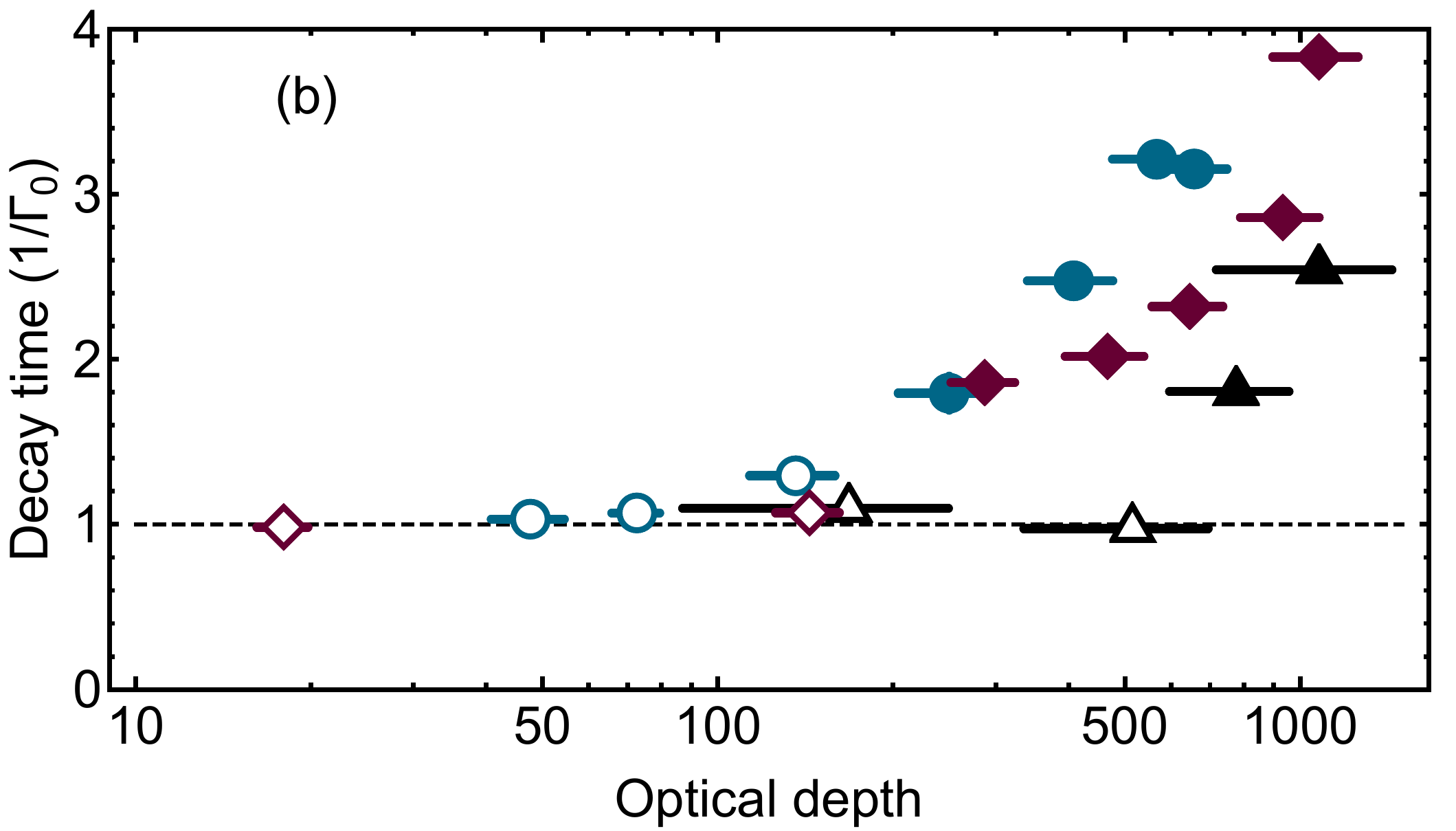} 
\includegraphics[width=\linewidth]{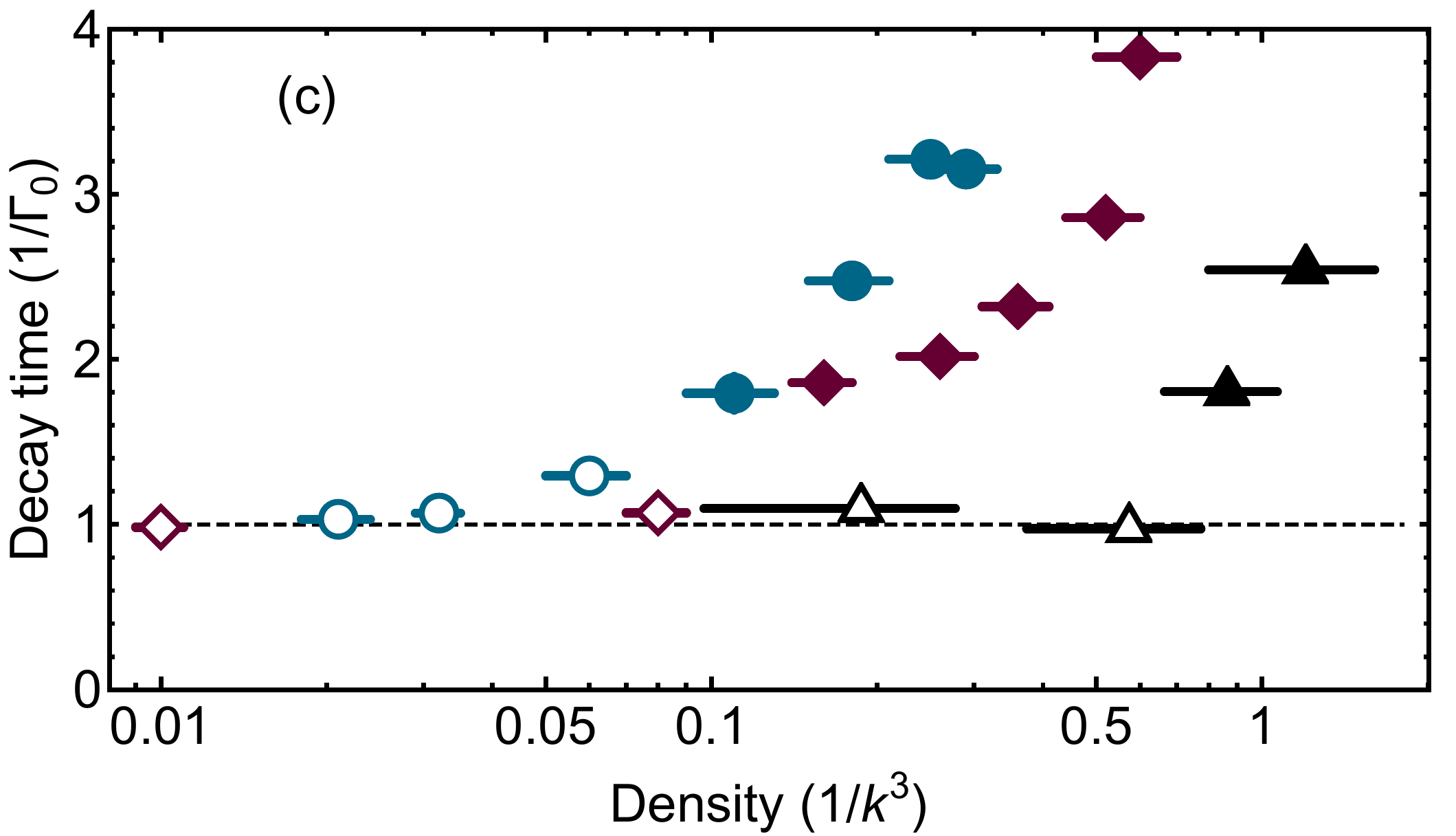} 
\caption{Decay times in units of the single atom lifetime $\Gamma_0^{-1}$ 
evaluated by fitting the traces with a sum of two exponential decays (filled symbols) 
or with a single one (empty symbols), as explained in the text.
The three different data sets are obtained in three different geometries 
giving cloud Gaussian sizes ($l_r, l_x$): 
($\SI{0.7}\lambda$, $\SI{7.7}\lambda$) (circles), 
($\SI{0.5}\lambda$, $\SI{6.0}\lambda$) (diamonds), 
and $(\SI{0.4}\lambda$, $\SI{2.9}\lambda$) (squares). 
(a) Experimental data as a function of the atom number.  
(b) and (c), same data plotted as a function of the optical depth $b_0$ 
and of the central cloud density $rho_0$. 
For all the measurements the saturation parameter was $s=I/I_{\rm sat}\simeq27$. Error bars on the decay time (standard errors from the fit) are smaller than the marker size.}
\label{fig2}
\end{figure}

To reveal the characteristic scaling of subradiance in our regime, 
we investigate the crossover between the low atom number regime, 
where the system behaves as an ensemble of non-interacting emitters, 
to the large atom number one. To adjust the atom number $N$, we 
release the atoms from the trap and recapture them after a variable time prior to sanding the burst of excitation pulses. 
This technique allows us to reduce the atom number by a factor more than 
10 with negligible heating, thus leaving the cloud sizes unchanged. 
To analyze the experimental data, we fit the decay with a phenomenological 
model using either a single exponential decay or the sum of two decays 
with different characteristic times. The fitting function is decided based on a $\chi^2$ criterion (see appendix\,\ref{app:fit} for more details). The decay time is extracted from data averaged over tens of thousands of realizations and thus corresponds to an average subradiant behaviour.\par 

We report in Fig.\,\ref{fig2}(a) the results of this analysis. We observe that as $N$ grows, 
a clear subradiant tail appears, and that the characteristic decay time is an increasing function of $N$. 
To determine the parameter that governs the cooperativity, 
we acquire three different data sets shown as different symbols in 
Fig.\,\ref{fig2}. They are taken in three different trapping geometries 
leading to different cloud sizes (appendix\,\ref{app:geoms}).
We plot the same data in Fig.\,\ref{fig2}(b) as a function of the 
optical depth along  $x$,
$b_0= \rho_0 l_x \sqrt{2\pi}\sigma_{\rm sc}$, 
and in Fig.\,\ref{fig2}(c) as a function of the peak density $\rho_0=N/[(2\pi)^{3/2}l_x  l_r^2]$ ($l_{x,r}$ are Gaussian sizes).
We observe that the data nearly collapse on a single curve when plotted as a 
function of atom number rather than as a function of optical depth. Furthermore, the decay times 
cannot be described either by the cloud density only, which is a local quantity. This is expected given the long-range character of the dipole-dipole interaction. We verify that the amplitude of the subradiant decay measured as the relative area in the tail (see appendix\,\ref{app:TR}) is also governed by the atom number only. Therefore the two parameters describing subradiance namely lifetime and amplitude are solely governed by $N$, which is the cooperativity parameter for this regime as explained in the introduction. This behaviour distinguishes our results 
from the ones in dilute systems where the cooperativity parameter is $b_0$ \cite{guerin2017light}, and indicates that we are approaching the Dicke limit. 
The imperfect collapse of the experimental data in Fig.\,\ref{fig2}(a) 
might be due to the fact that the system size along $x$ is still larger than 
the excitation wavelength. \par  

\section{Study of multiply-excited subradiant states}\label{sec:intensity}

%


Subradiance corresponding to the presence of long-lived excitations, a natural application 
would be to store light in an atomic medium. Storing multi-photon states would require long-lived multiply-excited states. 
Therefore understanding the nature of these excitations is a prerequisite for the application of multiple excitation storage \cite{cipris2020subradiance}. Here we investigate this question experimentally by varying the intensity of the excitation laser.  

\begin{figure}
\includegraphics[width=\linewidth]{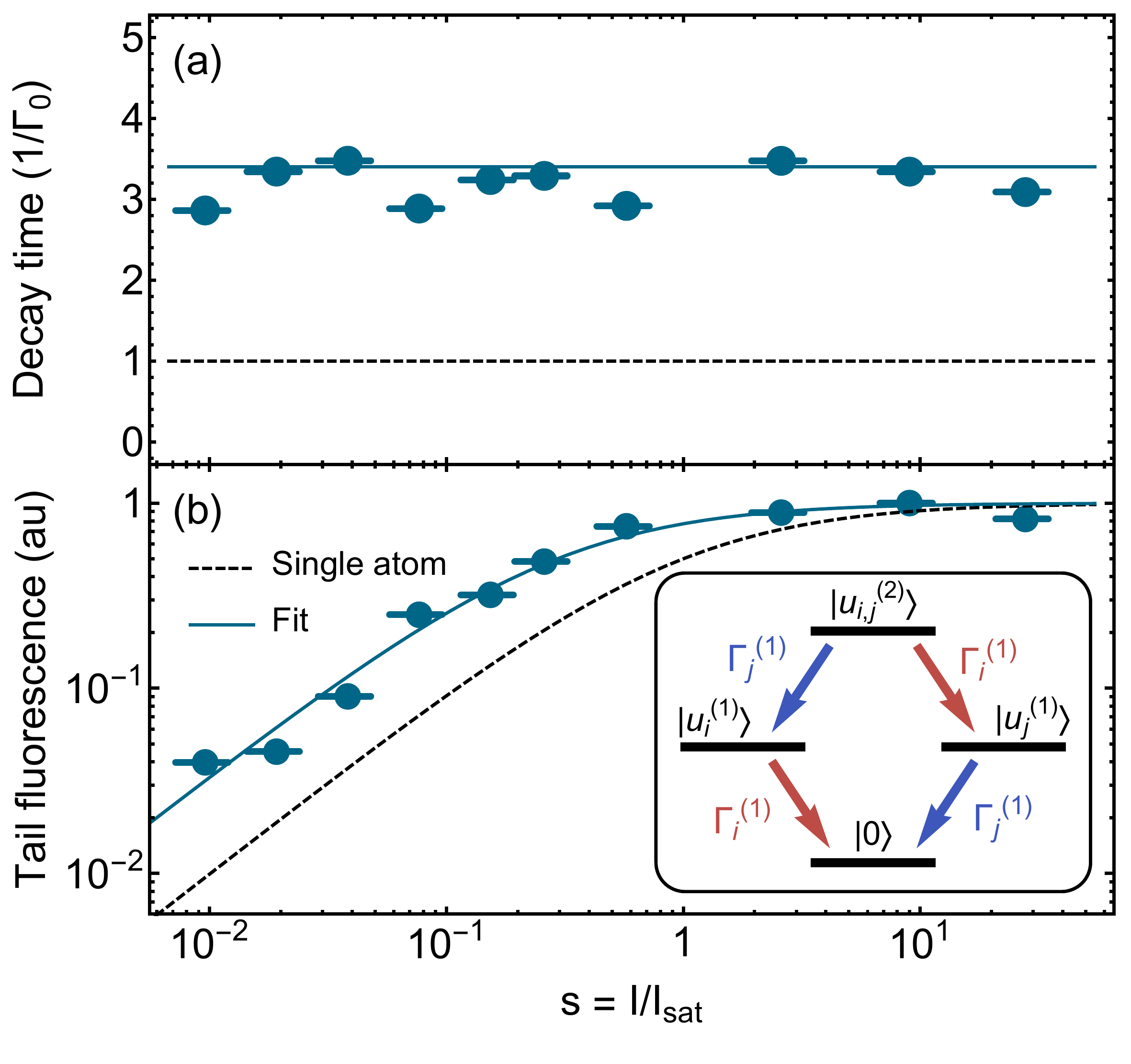} 
\caption{(a) Measurement of the decay time 
of the subradiant tail (unit of $\Gamma_0^{-1}$ )
as a function of the saturation parameter $s$ of the excitation laser. 
Black dashed line: single atom decay time.  
(b) Total number of counts recorded in the subradiant tail (normalized to maximum value) 
as a function of $s$, together with a fit by a function $\alpha s/(1+\alpha s)$ (blue solid line). 
From the fit we extract a decay time $\tau=\alpha\,\Gamma_0^{-1}$, 
see text, which is shown as a solid blue line in (a). 
Black dashed line: single atom response obtained setting $\alpha =1 $ in the previous equation. Caption: Independent decay process of multiply-excited subradiant states (here exemplified with $n_{\rm exc}=2$). Singly-excited states decay independently at a rate given by decay rates of the single excitation subradiant eigenmodes.}
\label{fig3}
\end{figure}

Considering first the strong driving limit, the ensemble is prepared in a product 
state where each atom is in a mixture of the ground $|g\rangle$ and excited state 
$|e\rangle$, with density matrix 
$\rho=\frac{1}{2^N}\left(|e\rangle\langle e|+|g\rangle\langle g|\right)^{\otimes N}$. In general, this mixture comtains subradiant components, as was discussed in ref.\,\cite{cipris2020subradiance} for the case of two atoms. Reaching such a steady state during the excitation therefore leads to the initial excitation of long-lived subradiant states. However it does not preclude the further population of these states during the early decay following the switch-off of the laser excitation \cite{cipris2020subradiance,masson2020,henriet2019critical}.
For large $N$, it was shown \cite{asenjo2017exponential,Albrecht2019Subradiant,henriet2019critical,zhang2020theory} 
in the case of ordered 1D arrays, that subradiant states  containing $n_{\rm exc}>1$ excitations are built from a superposition of 
subradiant states of the {\it single} excitation manifold, 
which decay independently with their respective lifetime $\Gamma_n^{(1)}$, as exemplified for $n_{\rm exc}=2$ in the caption of Fig.\,\ref{fig3}. The signal resulting from the decay of an $n_{\rm exc}>1$ state is $\propto\sum_{n=1}^{n_{\rm exc}}\exp(-\Gamma^{(1)}_n\,t)$. The interest of this ansatz stems from the fact that the singly-excited states can be calculated via a model of classical coupled dipoles \cite{ruostekoski1997quantum}.
It is however an open question whether this simple picture holds in the disordered case studied here.
As we show below, our experimental findings support the picture of 
multiply-excited subradiant states constructed as a superposition of 
independent singly-excited subradiant states. \par


To  control the number of excitations in the system, 
we vary the saturation parameter 
$s=I/I_{\rm sat}$ between $s\simeq0.01$ and $s\simeq30$. 
For these measurements we work without 
compression of the trap and the atom number is fixed $\simeq$\,4500. 
All the time traces acquired in this section are reported in appendix\,\ref{app:curves}. 
We extract the subradiant lifetime using the  procedure  introduced earlier. 
In addition, we calculate the tail fluorescence by summing the signal for times larger than 
$t_0+\frac{4}{\Gamma_0}$, where $t_0$ marks the switch-off of the excitation pulse. 
We observe that the subradiant decay time is 
constant over three orders of magnitude of the excitation intensity [Fig.\,\ref{fig3}(a)] 
and that the tail fluorescence increases with excitation intensity before saturating at an 
intensity smaller than $I_{\rm sat}$ [Fig.\,\ref{fig3}(b)]. The constant lifetime 
suggests a first simple description of the data as the excitation of a single mode. 
We thus use a single-mode approximation to describe the subradiant tail population 
with the following expression assuming a 
saturation behaviour similar to that of a single atom: $\propto\alpha s/(1+\alpha s)$, with $\alpha$ the 
fit parameter. We report the result of this fit as the solid line in Fig.\,\ref{fig3}(b). 
From the extracted $\alpha=3.4(5)$ and assuming that the lifetime 
of a given mode $\tau'$ dictates its saturation intensity $I_{\rm sat}' \propto1/\tau'$ 
as for a single atom \cite{GAF}, we obtain $\tau'=3.4/\Gamma_0$, 
represented as a solid line in Fig.\,\ref{fig3}(a). This result agrees remarkably
well with the direct measurement of the decay rate. Thus the single mode approximation 
describes very well our data, seeming to confirm the validity of this approximation. However in the saturated regime the long-lived modes leading to the subradiant decay host up to $10\,\%$ of the total excitations (see appendix\,\ref{app:TR} Fig.\,\ref{figtTRN}), which for a fully-saturated 
(i.e.~1/2 excitation per atom) cloud of 5000 atoms means several hundreds of excitations. Despite this large number of excitations, the decay rate remains the same demonstrating that, in the subradiant tail, the rate at which excitations decay is independent of the density of excitations in the system. This finding  is consistent with multiply-excited states constructed from a large population of singly-excited subradiant states which decay independently. In the opposite case of strong interactions between singly-excited subradiant states, we would have observed an excitation density-dependent decay rate due to additional decay processes induced by interactions between excitations. Experimentally we observe average decay times of about $3/\Gamma_0$, in agreement with the result of classical coupled dipole calculations showing that the singly-excitation manifold contains a large population of modes in the range around $3/\Gamma_0$ [see Sec.\,\ref{sec:release}, Fig.\,\ref{fig4}(b)]. 



Finally, we come back to the description of the dynamics in terms of the mean-field model introduced in sec.\,\ref{sec:charac}. This model assumes a factorizable density matrix throughout the decay, and the coupling between atoms is induced by their dipole moment proportional 
to the coherence between ground and excited states $\rho_{\rm eg}$ \cite{GAF}. However for high saturation the atoms are prepared in an incoherent mixture of the ground and excited states, hence the coherence and thus the dipole vanish. Therefore the mean-field model predicts a decoupling of the atoms with one-another, which  
then decay independently with the single atom lifetime $1/\Gamma_0$. Our observations up to $s=250$, see appendix\,\ref{app:highI}, contradict this prediction showing that the density matrix of the system cannot be factorized throughout the decay, although it can be factorized initially.\par

\section{Release of the subradiant excitation}\label{sec:release}



In this final section, inspired by theoretical proposals \cite{facchinetti2016storing,PhysRevA.94.013803,asenjo2017exponential}, we perform a proof-of-principle demonstration of the on-demand release of the light stored in subradiant excitations. To do so, we apply a position-dependent detuning.  The resulting inhomogeneous broadening makes the interaction between atoms no longer resonant: 
the ensemble now consists of independent atoms efficiently radiating, 
thus releasing the subradiant excitation. 

\begin{figure}[htbp]
\centering
\includegraphics[width=\linewidth]{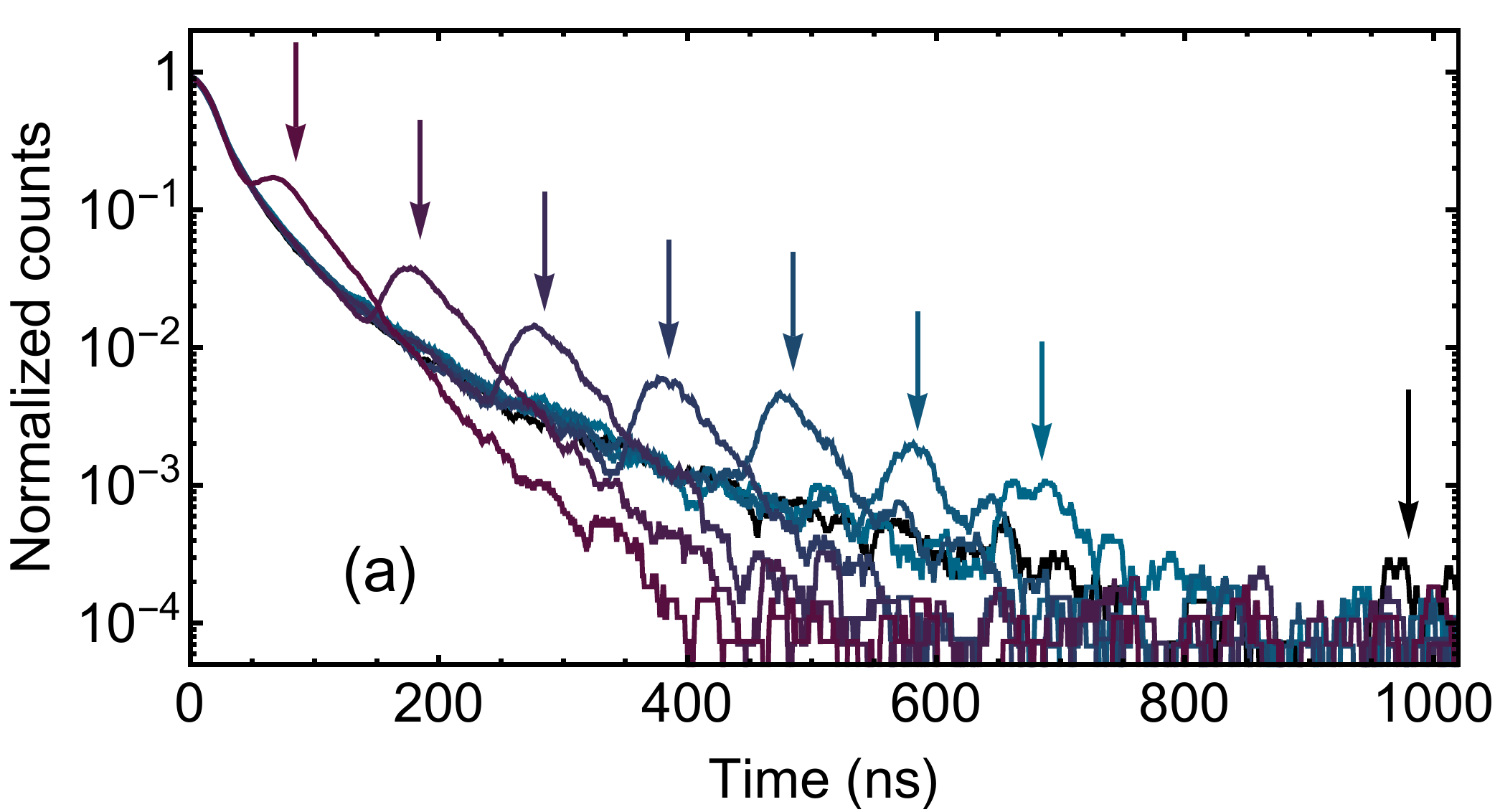}
\includegraphics[width=\linewidth]{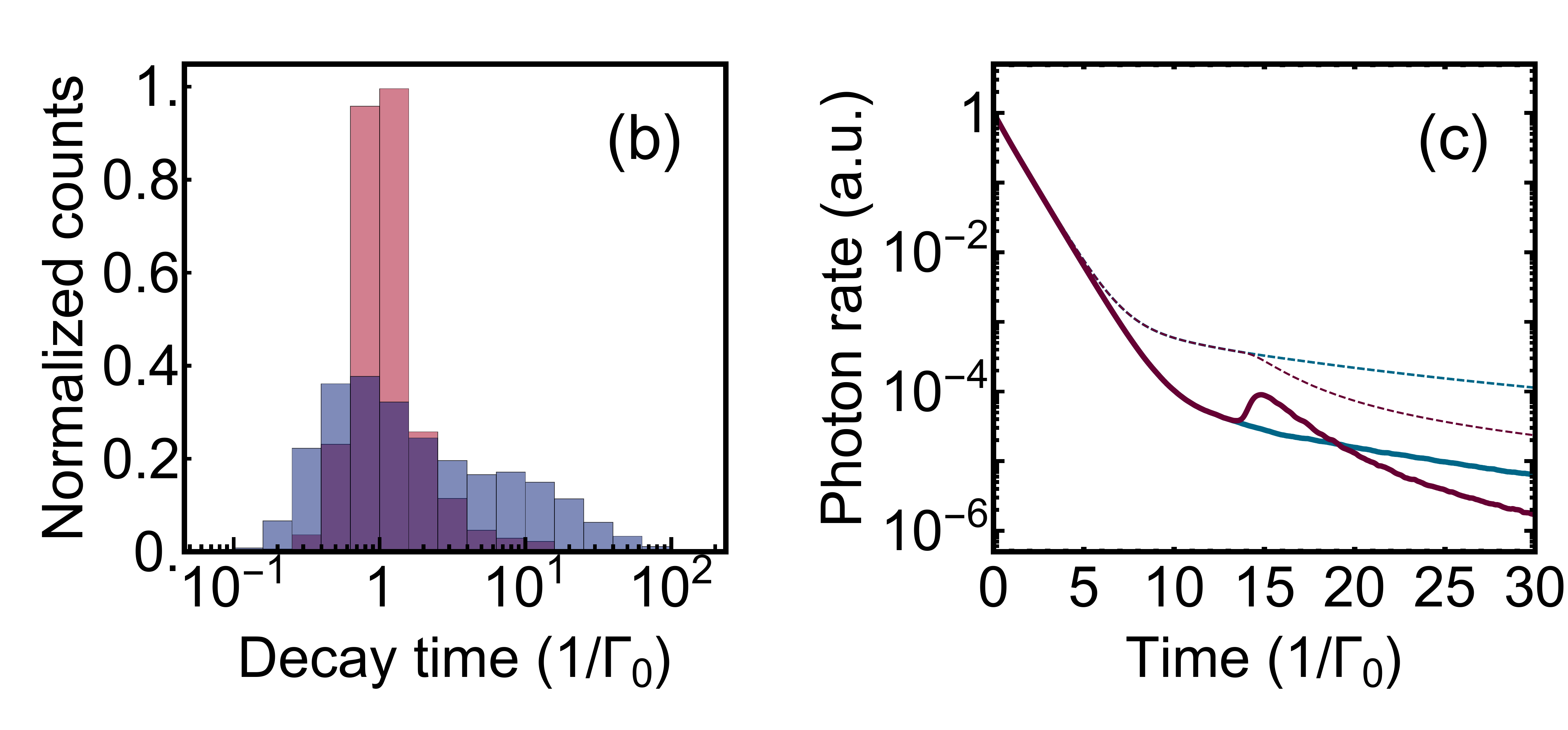}
\caption{ Release of the light stored in subradiant excitations. (a) Experimental realization, each  time trace represents an experimental sequence where the 
inhomogeneous broadening is applied at a different time, highlighted by the respective arrows. 
(b) Histogram of the decay times of the collective modes evaluated from the 
eigenvalues of the interaction matrix for 5000 atoms 
with the same density distribution as in the experiment, for 10 realizations. 
Red (blue) histograms are calculated with (without) inhomogeneous broadening. (c) Results of the mean-field non-linear coupled dipoles simulations. 
Dashed lines:  temporal evolution of the total excited state population $p(t)$.
Solid lines: intensity $-dp/dt$ emitted in a $4\pi$ solid angle. 
Purple lines: case where the inhomogeneous broadening 
is applied at  $t= 12.5/\Gamma_0$
Blue lines: no inhomogeneous broadening is applied.  
}
\label{fig4}
\end{figure}

To develop an intuition of how this protocol works, we first consider a toy model 
consisting of two coupled linear classical dipoles $d_1$ and $d_2$ with decay rate $\Gamma_0$ and separated by $\boldsymbol r_{12}$. 
The dynamics of the system is governed by the following equations (here $\hbar=1$):
\begin{equation}
\begin{pmatrix}
\dot{d}_1 \\
\dot{d}_2 
\end{pmatrix}=
\begin{pmatrix}
-\frac{\Gamma_0}{2} & i\,V(\boldsymbol r_{12}) \\
i\,V(\boldsymbol r_{12}) & i\delta-\frac{\Gamma_0}{2} 
\end{pmatrix}
\begin{pmatrix}
d_1 \\
d_2 
\end{pmatrix}
\label{eq0}
\end{equation}
where $\delta$ is the difference between the transition frequencies of the two 
atoms and $V(\boldsymbol r_{12})$ is the (complex) dipole-dipole interaction 
potential \cite{jackson2007classical}. 
This system has super- and a subradiant eigenmodes with decay rates $\Gamma_\pm=\Gamma_0\pm 2\,{\rm Im}{[V(\boldsymbol r_{12})]}$.
For  $\delta=0$ these modes are the symmetric and anti-symmetric combinations $\boldsymbol v_\pm\propto(1,\pm1)$. 
Hence, if at time $t=0$ the dipoles are prepared in $\boldsymbol v_-$, the turning on of
the inhomogeneous broadening ($\delta\neq0$) projects $\boldsymbol v_-$ onto 
the new basis provided by the eigenvectors of the matrix in Eq.\,\eqref{eq0}. 
In the limit $\delta\gg |V(r_{12})|$ these 
are the uncoupled dipoles $\boldsymbol v_1=(1,0)$ and $\boldsymbol v_2=(0,1)$, with identical decay rate 
$\Gamma_0$. The evolution of the two-atom dipole is then given by 
$\boldsymbol v(t)\propto e^{-\Gamma_0 t/2}(\boldsymbol v_1-e^{i\,\delta\, t}\boldsymbol v_2)$,
recovering the single atom decay rate of the radiated power $\boldsymbol v^2(t)\propto e^{-\Gamma_0 t}$. 
This toy model thus shows that placing the atoms out of resonance with one another allows the system to radiate again as an assembly of independent emitters.\par

We report in Fig.\,\ref{fig4}(a) the results of the experiment obtained applying an inhomogeneous broadening. To do so we turn on the trap light at different times (indicated by the vertical arrows) 
after the extinction of the excitation laser. The black curve is our reference for which the trap is turned on at long time ($\simeq\SI{1}{\micro \second}$). The far off-resonant light induces a position-dependent detuning described by:
\begin{equation}
\delta_i(x_i,y_i,z_i)=\frac{\delta_0}{1+x_i^2/x_R^2}\, \exp\left[-\frac{2(y_i^2+z_i^2)}{w_{\rm trap}^2(1+x_i^2/x_R^2)}\right]
\label{eq2}
\end{equation}
where $\delta_0$ is the light-shift induced by the trap. Experimentally $\delta_0\simeq32\,\Gamma_0$, 
$w_{\rm trap} =\SI{2.5}{\micro\meter}$ and $x_R = \pi w_{\rm trap}^2/\lambda_{\rm trap}$. Using the atomic density distribution in Eq.\,\eqref{eq2} for the $x_i,\,y_i,\,z_i$, 
the standard deviation of the detuning induced by the trap is about $4\,\Gamma_0$. At the turn on of this inhomogeneous broadening, we observe the emission of a pulse of light [Fig.\,\ref{fig4}(a)]. The presence of this pulse can be qualitatively understood using the toy model: when the inhomogeneous broadening is applied the atoms start to radiate at a rate faster than the subradiant one, thus the intensity of the emitted light is initially enhanced before decaying. Moreover we observe that the pulse is followed by a faster decay similar to the single-atom regime (fitting the experimental data after this pulse we obtain a decay rate of $1.3(1)\Gamma_0$
for all data sets).
The measurements have been performed with $\simeq$ 5000 atoms and with $s\simeq $ 27. 
We verified the collective origin of this effect by performing the same measurements
for low atom number, observing no pulse. Furthermore the strong suppression of subradiance obtained with a light-shift varying slowly in space demonstrates that the subradiant excitations are not stored in pairs of closeby atoms  
but rather delocalized over all the atoms of the cloud \cite{schilder2016polaritonic}. 
This is expected for a near-resonant excitation of the cloud, as delocalized excitations corresponds 
to states with small interaction frequency shifts.\par

To go deeper in the understanding of the observed behavior, we extend the toy model previously introduced to large ensembles. We obtain  the decay rates by evaluating the eigenvalues $\lambda_i$ 
of the interaction matrix for $N=5000$ classical coupled dipoles sampled using the experimental position-dependent detuning.
The associated modes are single-excitation modes, 
but as observed earlier they should give a qualitative description of the 
behaviour of the subradiant tail. 
The results are shown in Fig.\,\ref{fig4}(b). They indicate
that in the presence of the inhomogeneous broadening, the distribution of
decay times is much narrower than in free space [note the log scale in Fig.\,\ref{fig4}(b)].
In particular a significant fraction of modes with decay rates close 
to the observed subradiant one (between $\Gamma_0/3$ and $\Gamma_0/10$) 
is suppressed by the inhomogeneous broadening. We further performed numerical simulations of the dynamics for smaller atom numbers using the mean-field model 
already introduced in Section \ref{sec:charac}. The results are shown in Fig.\,\ref{fig4}(c) for a 
cloud with $N=200$ atoms at a density $\rho_0/k^3 = 0.3 $. 
The trap is turned on during the decay. 
We took into account the finite rise time of the trap beam of about  $\SI{25}{\nano\second}$.  
The simulation provides the evolution of the total population of 
the excited state $p(t)=\sum_{i=1}^N\rho_{\rm ee}^{(i)}(t)$ (dashed line). 
However, experimentally we measure the intensity of the light emitted by the cloud, which in the absence of a drive is proportional 
to $-dp(t)/dt$ (in a $4\pi$ solid angle) represented by the continuous lines of Fig.\,\ref{fig4}(c).
When the inhomogeneous broadening is applied, the population curve presents a change of slope, 
corresponding to a peak in the intensity, 
followed by a decay at a rate now close to the single atom one. These simulations confirm the interpretation of our experimental findings \par

\begin{figure}[htbp]
\centering
\includegraphics[width=\linewidth]{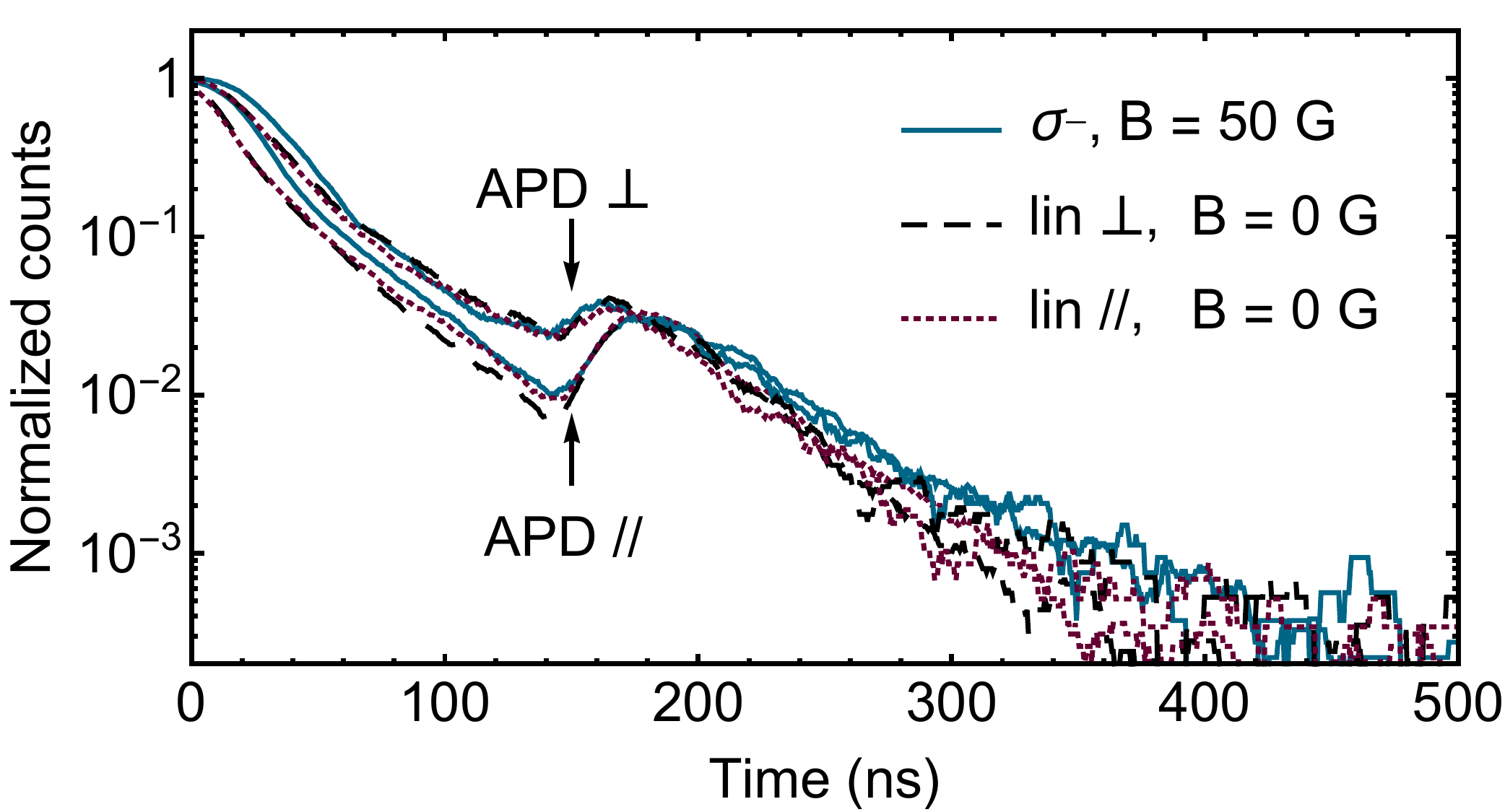}
\caption{Directionality of the emitted light pulse. Release of the subradiant excitation, observed with light collected axially (APD\,//) (bottom curves) and radially (APD$\,\perp$) (top curves). Experiment done with a linear polarization of the excitation light, 
either parallel (dotted lines) or perpendicular (dashed lines) to the cloud, and with a circular polarization in the presence of a 50\,G magnetic field alligned along the excitation direction (solid lines). }
\label{fig5}
\end{figure}

Finally, as shown in Fig.\,\ref{fig5}, we observe that the enhancement of 
the emission is stronger along the cloud axis. 
We have further observed by changing polarization and the internal structure of the atoms 
that no significant difference arises in the released pulse shape, Fig.\,\ref{fig5}. 
We leave for future work the investigation of how to control 
the directionality of the pulse which could allow for the tailoring of the angular emission pattern of the excitation stored in subradiance.\par


\section{Conclusion}
In this work, we investigated the subradiant decay of excitations stored in a dense 
cloud of atoms trapped in an optical tweezer, approaching the 
Dicke limit of a large number of atoms in a volume smaller than $\lambda^3$. 
We confirm that the cooperativity parameter is the atom number rather that the optical depth or the density. 
Moreover, by tuning the intensity of the excitation laser, we studied the nature of multiply-excited subradiant states and experimentally find that all the information is contained in the response of the system at low intensity where only singly-excited modes are populated. 
A quantitative theoretical description in our regime is however extremely challenging 
and new models describing this interacting dissipative many-body system need to be developed.
Finally, by applying an inhomogeneous broadening, we were able to release 
the subradiant excitation stored in the cloud in the form of a pulse of light. 
Our experiment thus represents a proof-of-principle of the temporal  control of 
subradiance in an atomic medium, a prerequisite for its use as a light storage medium. 
It was made possible by the small size of the cloud, which permitted the use of a relatively low 
power of the control light to apply a significant detuning between atoms. In the future it could be applied to 
systems featuring tailored subradiant modes, such as structured atomic systems  
\cite{asenjo2017exponential, bettles2016cooperative, bettles2016enhanced, shahmoon2017cooperative}, 
in which these modes could be targeted by a local addressing.\par

\begin{acknowledgements}
	We thank Ana Asenjo-Garcia, Stuart Masson, Luis Orozco, Robin Kaiser and William Guerin for discussions. This project has received
funding from the European Union’s Horizon 2020 research
and innovation program under Grant Agreement
No. 817482 (PASQuanS) and by the R\'egion \^Ile-de-France
in the framework of DIM SIRTEQ (project DSHAPE). A.\,G.~is supported by the D\'el\'egation G\'en\'erale de l’Armement Fellowship No.\,2018.60.0027.
\end{acknowledgements}
\bibliography{biblio}
\clearpage

\newpage

\setcounter{figure}{0}
\renewcommand\thefigure{A\arabic{figure}} 
\setcounter{equation}{0}
\renewcommand\theequation{A\arabic{equation}} 
\appendix
\section{Non-linear coupled-dipoles model}\label{app:NLCD}
In this section we provide some details about the mean-field non-linear coupled-dipoles model used in the main text. A derivation of the main equations can be found elsewhere \cite{do2020collective,Glicenstein2020}. We consider an ensemble of $N$ two-level atoms at position $\boldsymbol r_n$ coupled to light by a transition with dipole element $d_0$, detuning $\Delta$ and natural broadening $\Gamma_0$. Assuming that the density matrix can be factorized, $\rho=\otimes_n \rho_n$, the dynamics is governed by a system of coupled OBEs: 
\begin{equation}
\frac{d\rho_{\rm ee, n}}{dt}= -\Gamma_0\rho_{{\rm ee}, n}+i\rho_{{\rm ge}, n}\frac{\Omega_n}{2}-i\rho_{{\rm eg}, n}\frac{\Omega_n^*}{2}
\label{NLCD1}
\end{equation}
\begin{equation}
\frac{d\rho_{{\rm eg}, n}}{dt}=\big{(} i\Delta - \frac{\Gamma_0}{2}\big{)}\rho_{{\rm eg}, n}-i\Omega_n(\rho_{{\rm ee}, n}-\rho_{{\rm gg}, n})
\label{NLCD2}
\end{equation}
where $\rho_{\rm \alpha\beta, n}$ represents the elements of the density matrix of the $n$th atom. The driving Rabi frequency $\Omega_n$ experienced by atom $n$ is the sum of the laser driving and the field scattered by other dipoles: 
\begin{equation}
\begin{aligned}
\Omega_n=\Omega_{\rm laser}(\boldsymbol r_n) + i\, \sum_{m\neq n} \rho_{{\rm eg},m} V(\boldsymbol r_n-\boldsymbol r_m)\\
 V(\boldsymbol r) = \frac{3}{2i}  \frac{e^{ikr}}{kr}\bigg{[} p(\boldsymbol \hat{r})+ q(\boldsymbol \hat{r}) \bigg{(}\frac{i}{kr} -\frac{1}{(kr)^2}  \bigg{)}\bigg{]}
\end{aligned}
\end{equation}
where $p(\boldsymbol \hat{r})$ and $q(\boldsymbol \hat{r})$ depend on the polarization of the driving field \cite{akkermans2008photon}.\par
In the strong driving limit, $\rho_{{\rm eg},n}$ vanishes. Moreover, when the driving is turned off ($\Omega_{\rm laser}=0$), at any subsequent time $\rho_{{\rm eg},n}=0$, and the equations are decoupled leading to the single atom decay. 
%

\section{Dependence of subradiance on detuning}\label{app:det}
\begin{figure}[h!]
\includegraphics[width=\linewidth]{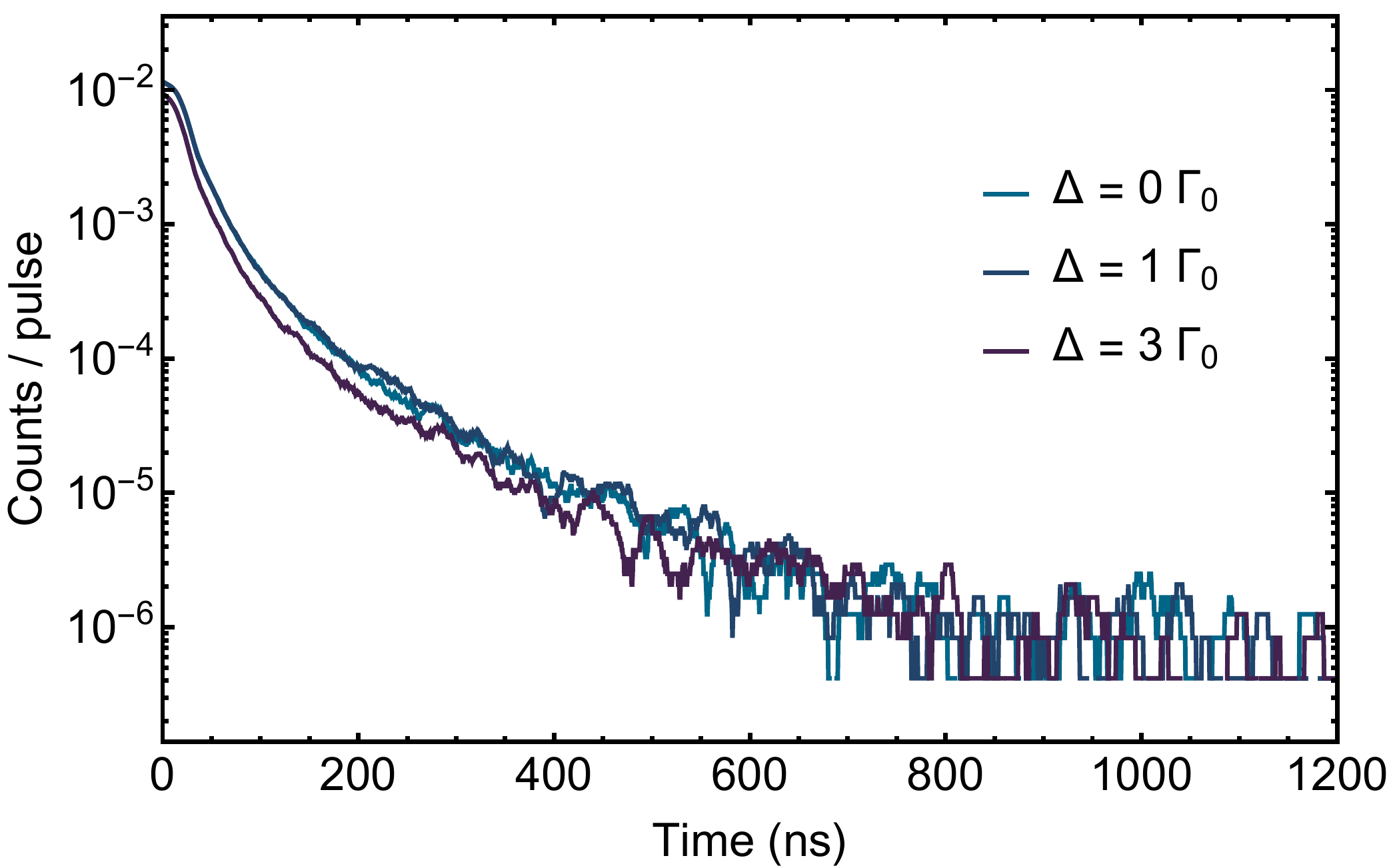} 
\caption{Photon count traces acquired as a function of the excitation frequency. The experimental data have been acquired in the shallowest trapping geometry described in the main text with $N \simeq 4500$. The temporal traces have been divided by the number of pulses of light used, they thus represent the mean number of photons collected during one pulse of resonant light.}
\label{figSI1}
\end{figure}

All the measurements reported in the main text have been performed on resonance. To demonstrate that our observations cannot be explained by radiation trapping which depends strongly on detuning \cite{guerin2016subradiance}, we have also acquired three different data sets at three different detunings of the driving laser. In Fig.\,\ref{figSI1} we report these measurements performed at resonance $\Delta/\Gamma_0=0$,  $\Delta/\Gamma_0=1$  and $\Delta/\Gamma_0=3$. At $\Delta/\Gamma_0=3$ we observe a small reduction in fluorescence at large detunings even at this intensity. However, the tail of each distribution behaves very similarly, independently of the detuning. 

\section{Fitting procedure to extract the subradiant lifetime}\label{app:fit}
\begin{figure}[h!]
\includegraphics[width=0.85\linewidth]{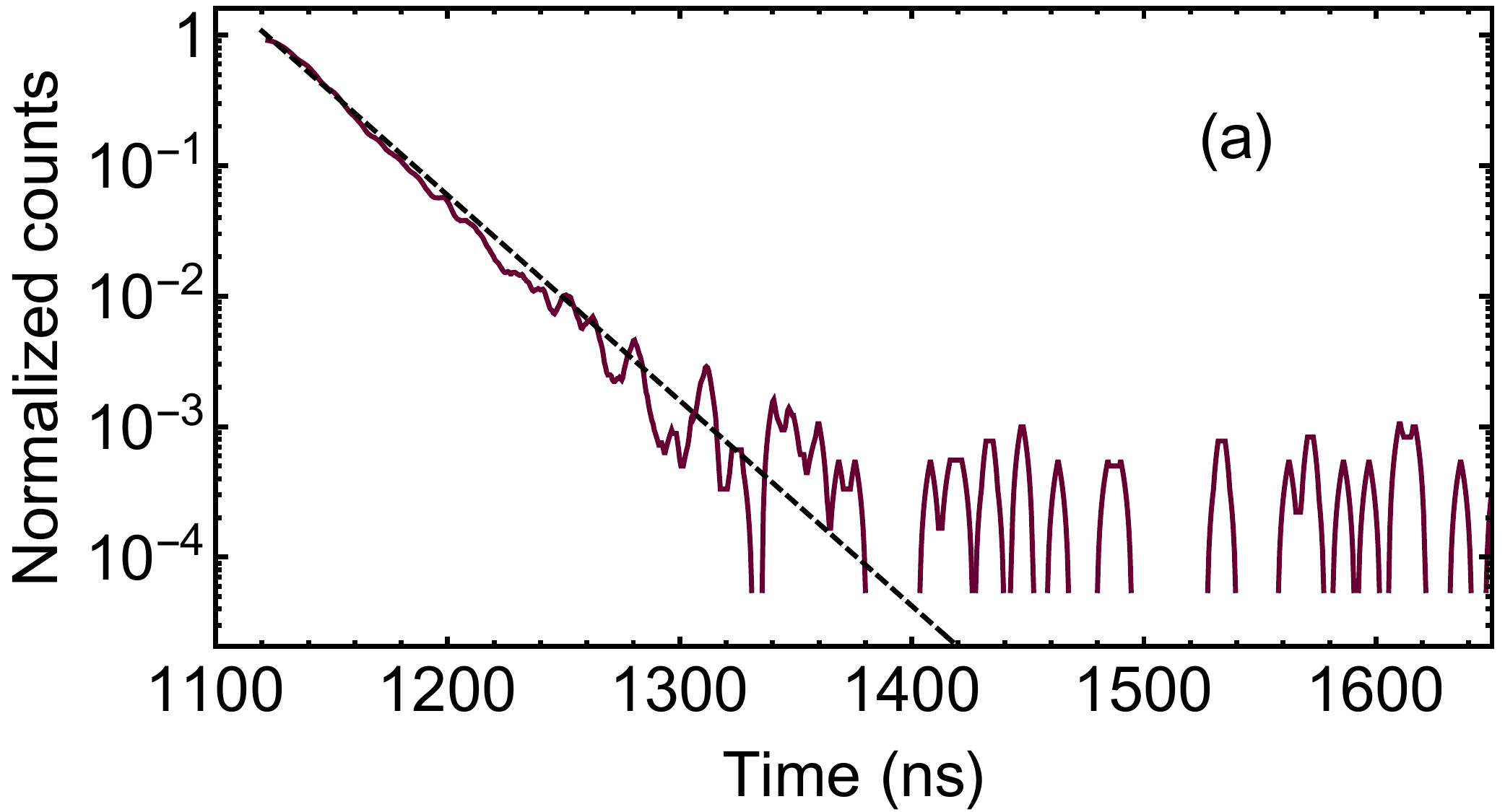} 
\includegraphics[width=0.85\linewidth]{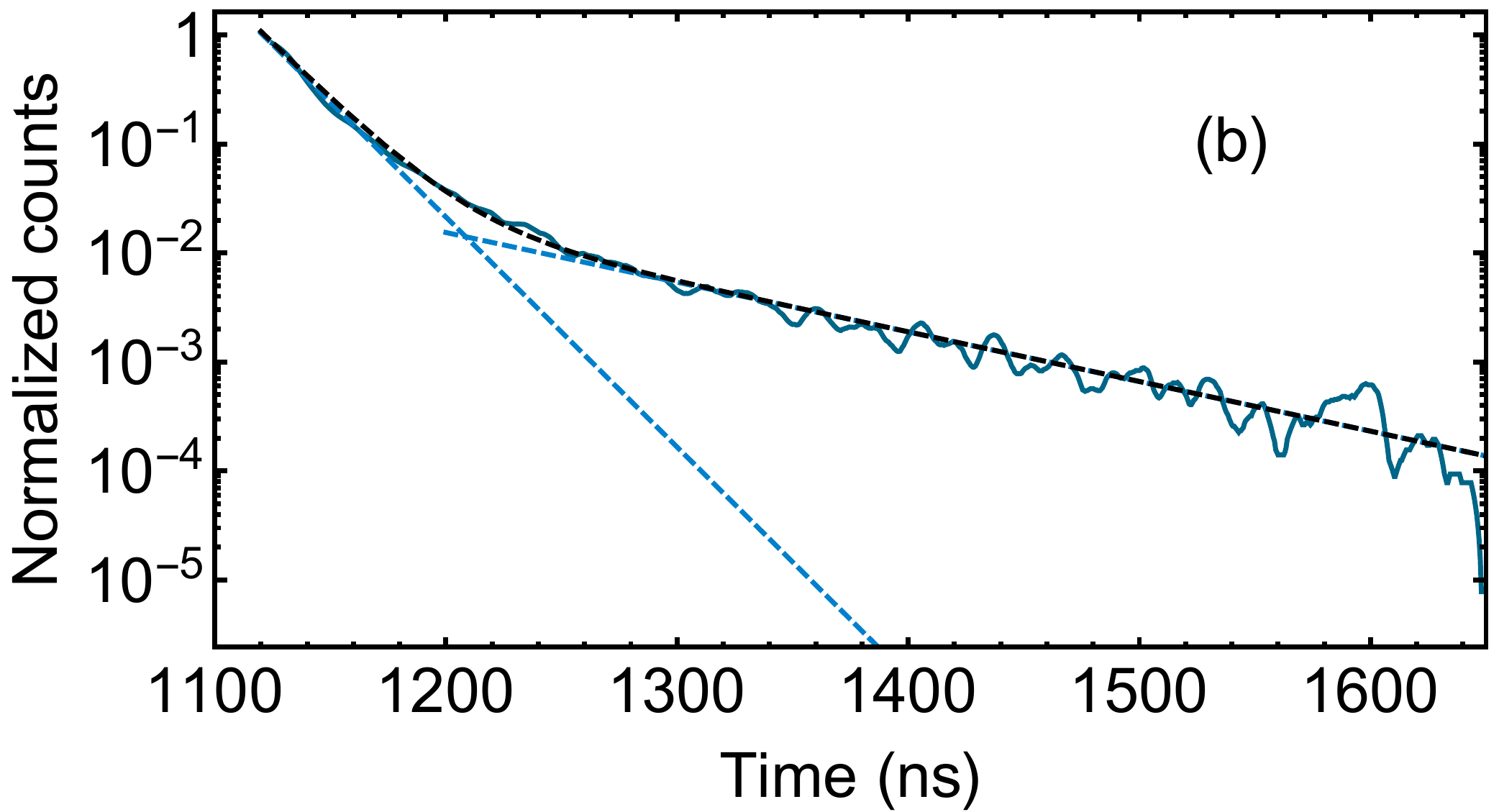} 
\includegraphics[width=0.85\linewidth]{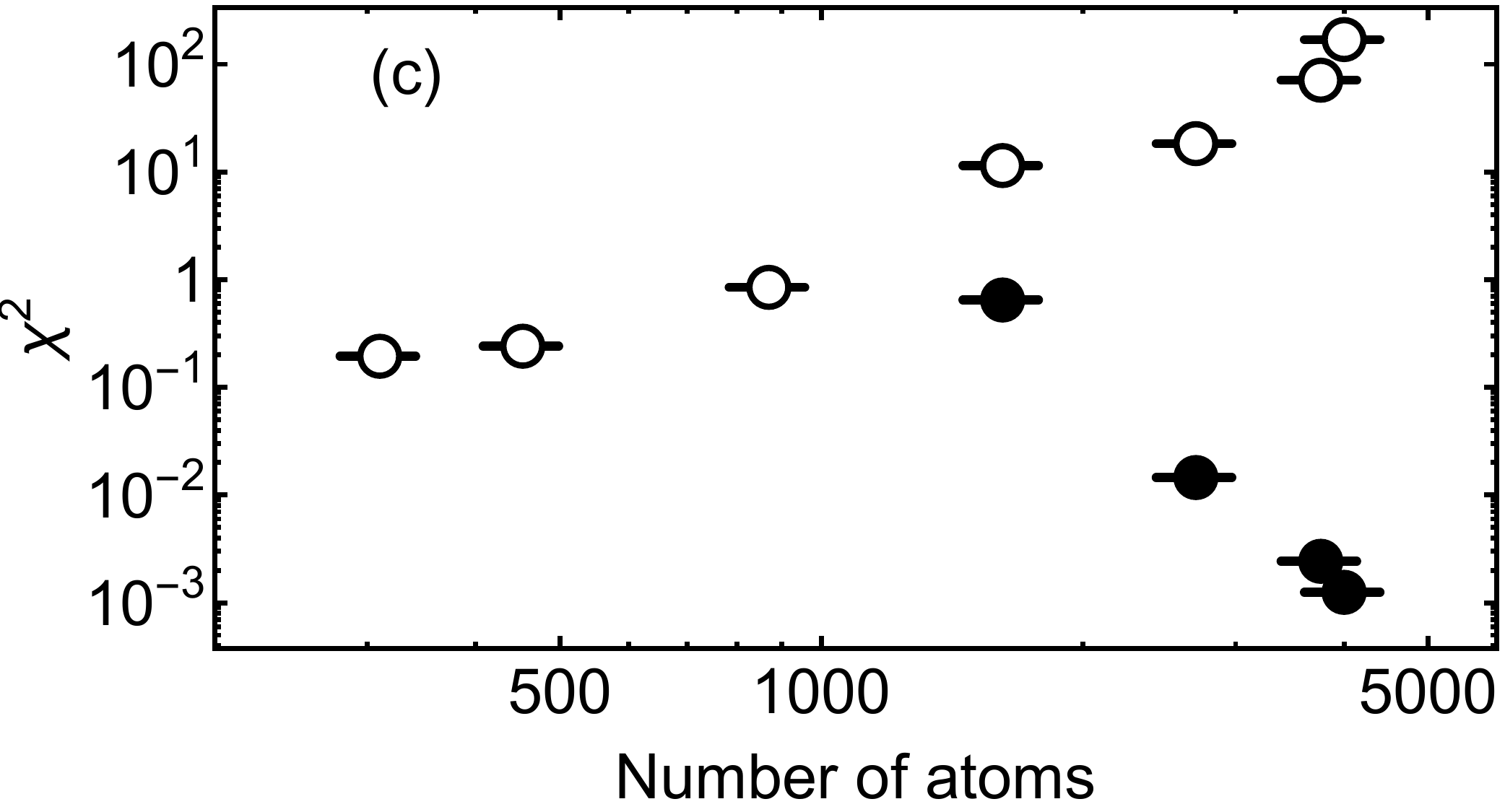} 
\caption{(a) Solid line: example of decay in the small atom number regime ($N\simeq300$) shown with the fit using a single exponential decay (dashed line). (b) Solid line: decay in the large atom number regime ($N\simeq4000$), dashed line: phenomenological fit with the sum of two exponential decays. The dotted lines represent the two different decay rates, i.e., super and subradiance. The traces in (a) and (b) have been normalized to the steady state. (c) Values of $\chi^2$ obtained with the double-exponential (filled circles) or a single exponential decay (empty markers) versus atom number.}
\label{figSI2}
\end{figure}

In this section we provide more details about the fitting procedure used to analyze the experimental data. We report in  Fig.\,\ref{figSI2}(a) an example of a fluorescence trace acquired in the small atom number regime together with a fit that makes use of a single exponential decay. We fit all the data in this way and for every measurements we evaluate the reduced $\chi^2$:
\begin{equation}
	\chi^2=\frac{1}{K_{\rm bin}}\sum_{i=1}^{K_{\rm bin}}\frac{(N_i^{\rm exp}-N^{\rm the}(t_i))^2}{N^{\rm the}(t_i)},
\end{equation}
where $K_{\rm bin}$ is the number of time bins in the data set ($\SI{500}{\pico \second}$ bins), $N_i^{\exp}$ is the recorded number of counts in the bin centered on time $t_i$ and $N^{\rm the}(t_i)$ is the value of the fitting function. This definition of $\chi^2$ assumes poissonian statistics for the recorded counts in each bin. \\

 In  Fig.\,\ref{figSI2}(c) we report the values of $\chi^2$ as a function of the atom number, for the shallowest trapping geometry described in the main text (blue points in Fig.\,\ref{fig2}). As the atom number in the cloud becomes larger, the system exhibits super- and subradiance and the fit with a single exponential decay is no more able to describe the entire observed dynamics as revealed by an increase in $\chi^2$. When it becomes larger than 1 we use a different phenomenological model to describe the decay, fitting with the sum of two exponential decays [see  Fig.\,\ref{figSI2}(b)]. The agreement between this second model and the experimental data is much better in the large atom number regime, and consequently $\chi^2$ is smaller as highlighted by the filled points in  Fig.\,\ref{figSI2}(c). The very small values of $\chi^2$ comes from the large uncertainties resulting from the small number of counts in the tail.\par
 
\section{The three different trapping geometries}\label{app:geoms}
In section \ref{sec:charac}, we employ three different trapping geometries that allow us to obtain three different cloud dimension. In the first one, the trap is the one used to load the atoms from the MOT (Gaussian sizes $l_r\simeq\SI{0.7}\lambda$, $\l_x\simeq\SI{7.7}\lambda$ blue circles). In the second one we increase the power of the trapping beam by $\simeq$ 60$\%$ gaining a factor $\simeq\si{2}$ in density without losing atoms ($\l_r\simeq\SI{0.5}\lambda$, $\l_x\simeq\SI{6.0}\lambda$, purple diamonds). In the third case, we reduce the beam waist down to $\SI{1.8}{\micro\meter}$. The atom number after the compression is lower ($N\simeq$1500) but we reach a high density $\rho_0/k^3\simeq$1.5 owing to the reduced trapping volume ($\l_r\simeq\SI{0.4}\lambda$, $\l_x\simeq\SI{2.9}\lambda$ black squares), \cite{Setup}. A complete set of time traces can be found below. For every point we measure the atom number and the temperature.\par

\section{Effect of polarization and optical pumping}\label{app:levStruc}
Unless otherwise specified, the measurements reported in this work are performed with a linearly polarized excitation light without any direct optical pumping (OP). We did not observe that the polarization direction impacts in any way the observed subradiance. However, there could be an effect due to the fact that the linear polarization configuration does not correspond to a two-level configuration.\par 
In order to check if this strongly modifies the collective behavior hosted in the tail of the fluorescence signal, we performed one experiment in a situation much closer to a two-level configuration: a circularly-polarized ($\sigma^-$) excitation, together with a $\SI{20}{G}$ magnetic field along the excitation propagation axis. We perform hyperfine and Zeeman optical pumping with the same polarization as the excitation light to place ourselves in a closed two-level system (between $F=2$, $m_F=-2$, $F'=3$, $m_F=-3$). The light scattered by an atom on a nearby one might contain other polarization components and drive other transitions out of the 2-level system. However the detuning between the $\sigma^-$ and $\pi$ polarization transitions (closest in detuning) is about $5\,\Gamma_0$ preventing the rescattering of this polarization by nearby atoms.\par

\begin{figure}[h!]
\includegraphics[width=\linewidth]{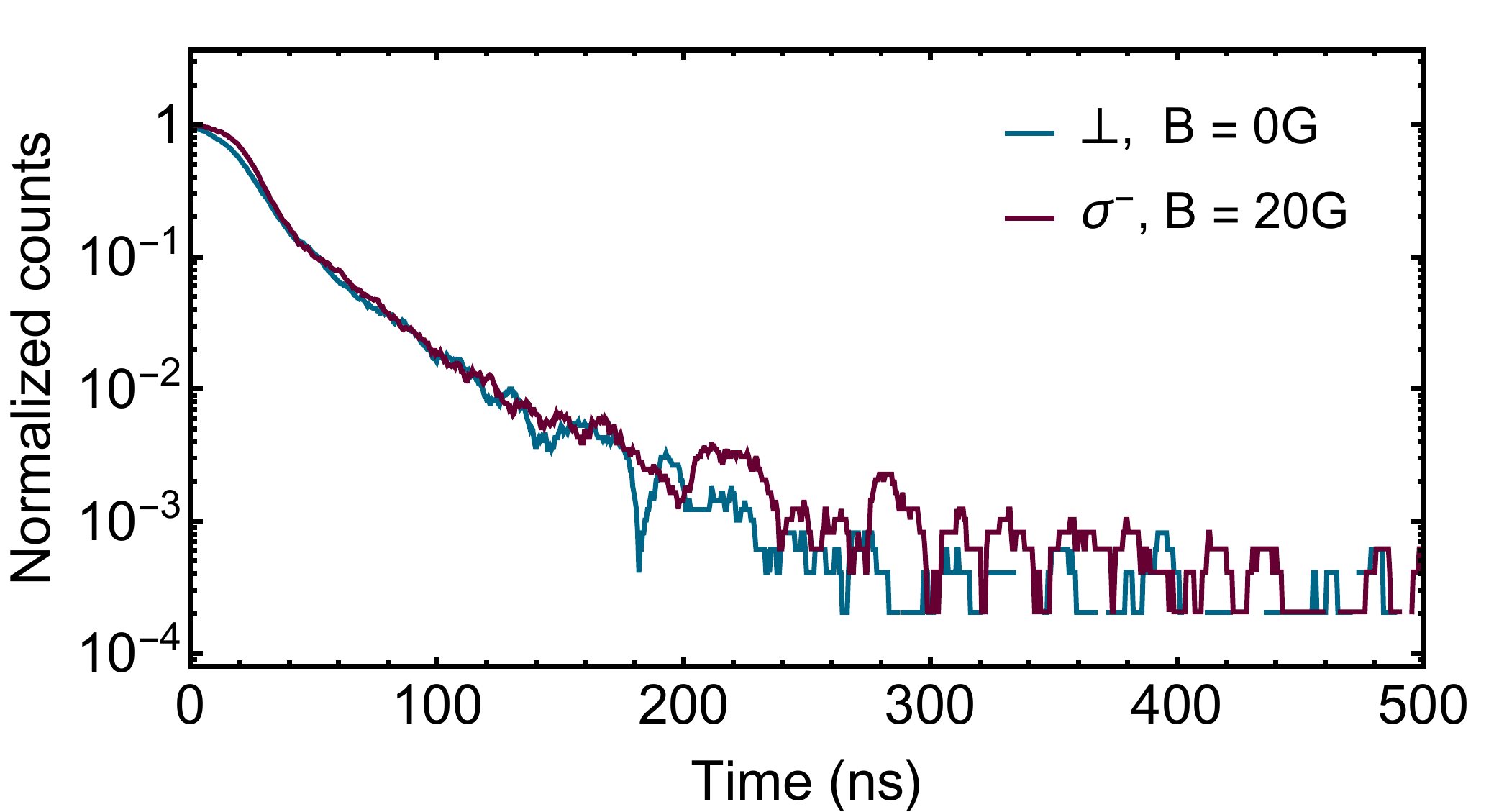} 
\caption{Photon count decay with a linear polarization, no magnetic field and no optical pumping i.e.~multi-level situation (blue). The same in the two-level case: $\SI{20}{G}$ magnetic field, $\sigma^-$ polarization and prior optical pumping (purple). The two traces have been normalized to the steady state, measurements for $N\simeq 4500$ in the first trapping geometry (see appendix\,\ref{app:geoms}).}
\label{figSI3}
\end{figure}

In  Fig.\,\ref{figSI3} we report in blue an acquisition done with linear polarization and without optical pumping, and in purple that with optical pumping, circular polarization and magnetic field. We observe that the tails of the two traces are similar, within our dynamic range of observation. We conclude from this that the internal structure does not seem to affect strongly the results on subradiance reported in this work and our conclusions. 

\section{Relative population of the long-lived states}\label{app:TR}
Another parameter that we use to quantify subradiance is the relative fluorescence observed in the long-lived tail with respect to the total fluorescence recorded after the switch off of the drive. We refer to this quantity as the tail ratio. It is defined as:
\begin{equation}
{\rm TR}=\frac{\int_{\frac{4}{\Gamma_0}}^\infty I(t)dt}{\int_{0}^\infty I(t)dt}\label{eq3}
\end{equation}
where $I(t)$ represents the time-resolved fluorescence emitted by the atomic cloud where $t=0$ is the switch off time of the excitation laser. This parameter estimates the fraction of excitation still hosted in the system $4/\Gamma_0\simeq\SI{100}{\nano\second}$ after switching off the excitation, compared to the whole decay. In the single-atom limit, this parameter is equal to $e^{-4}\simeq 2 \%$ by definition.\par
\begin{figure}
\includegraphics[width=0.85\linewidth]{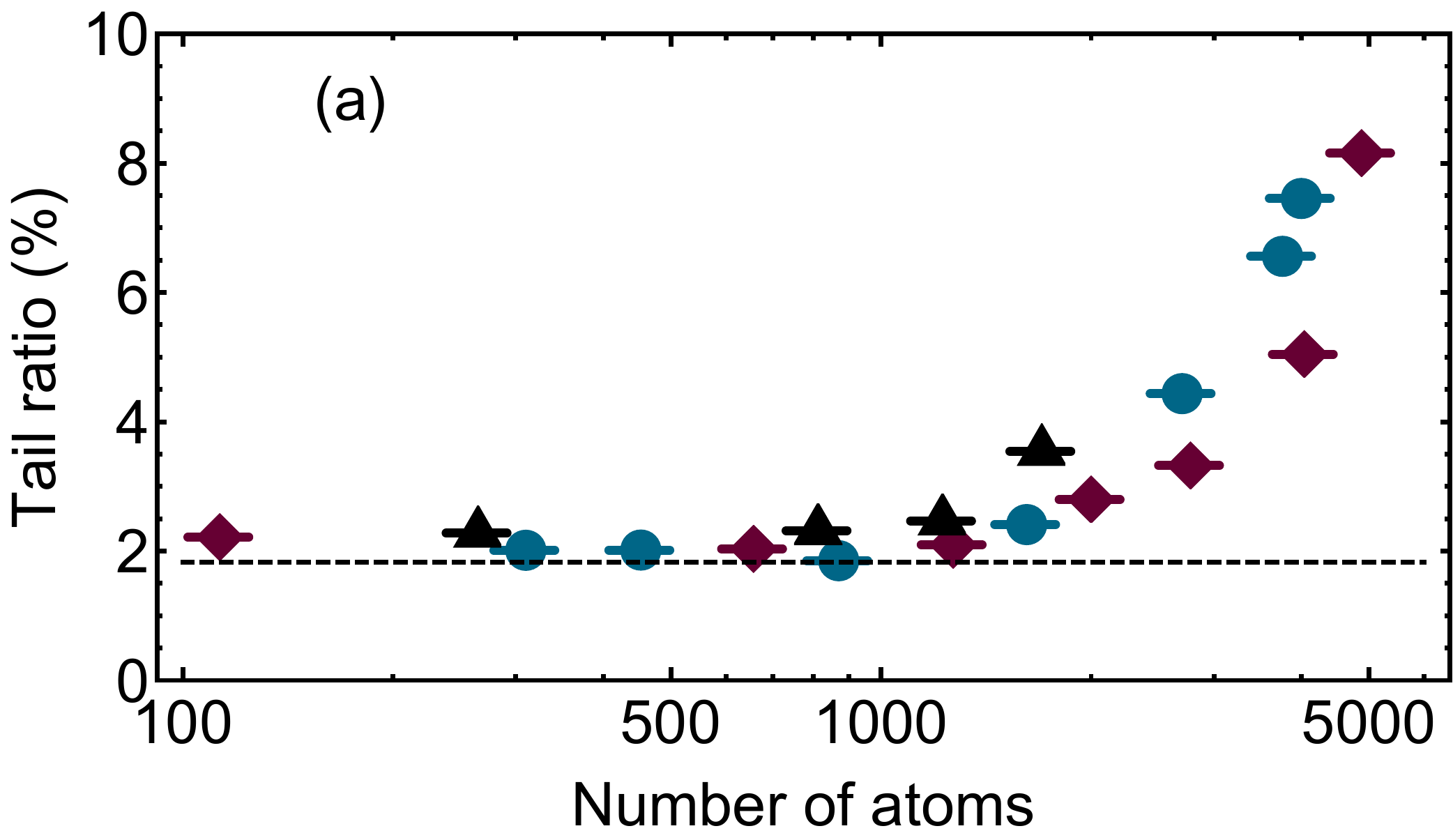} 
\includegraphics[width=0.85\linewidth]{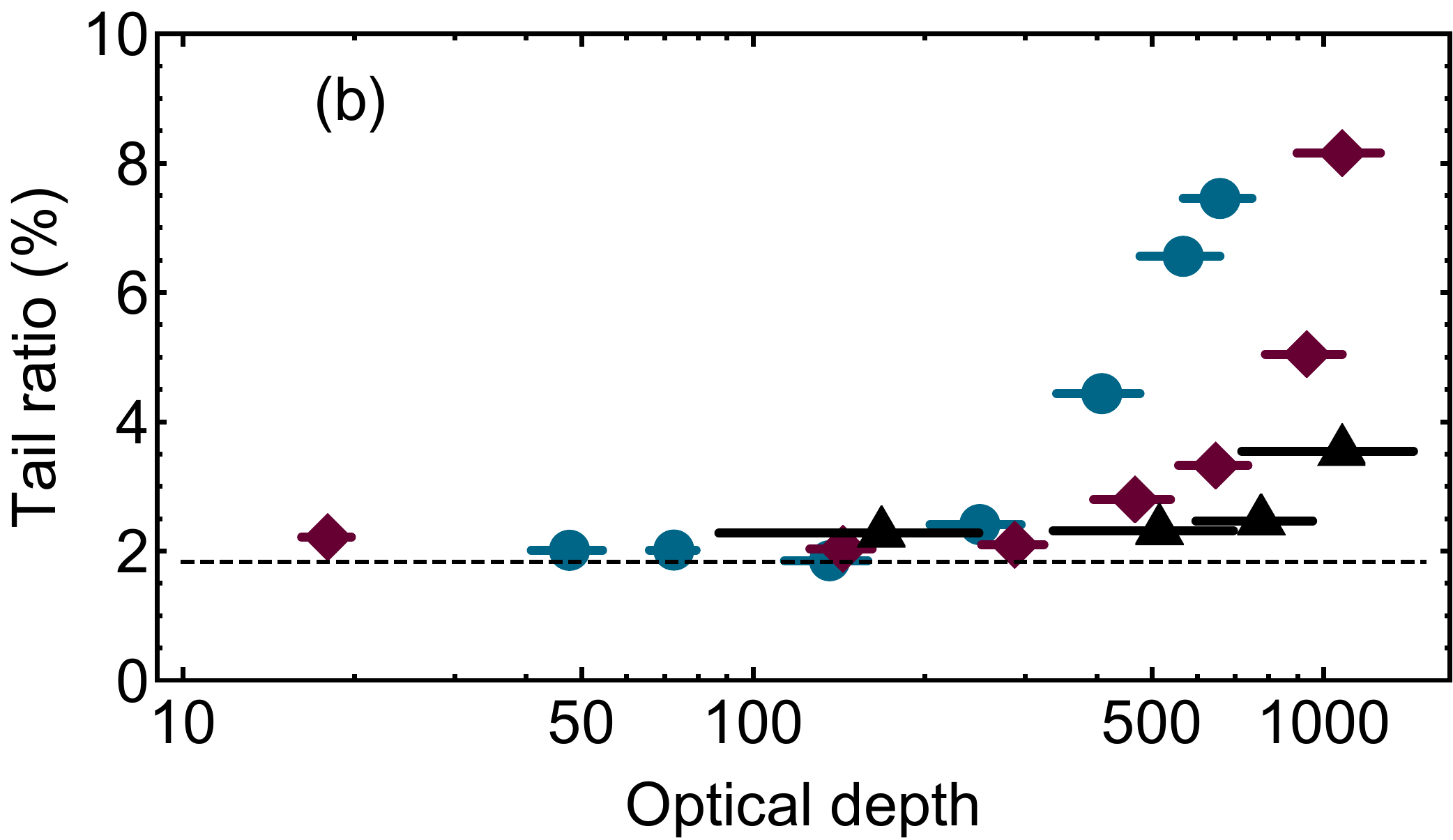} 
\includegraphics[width=0.85\linewidth]{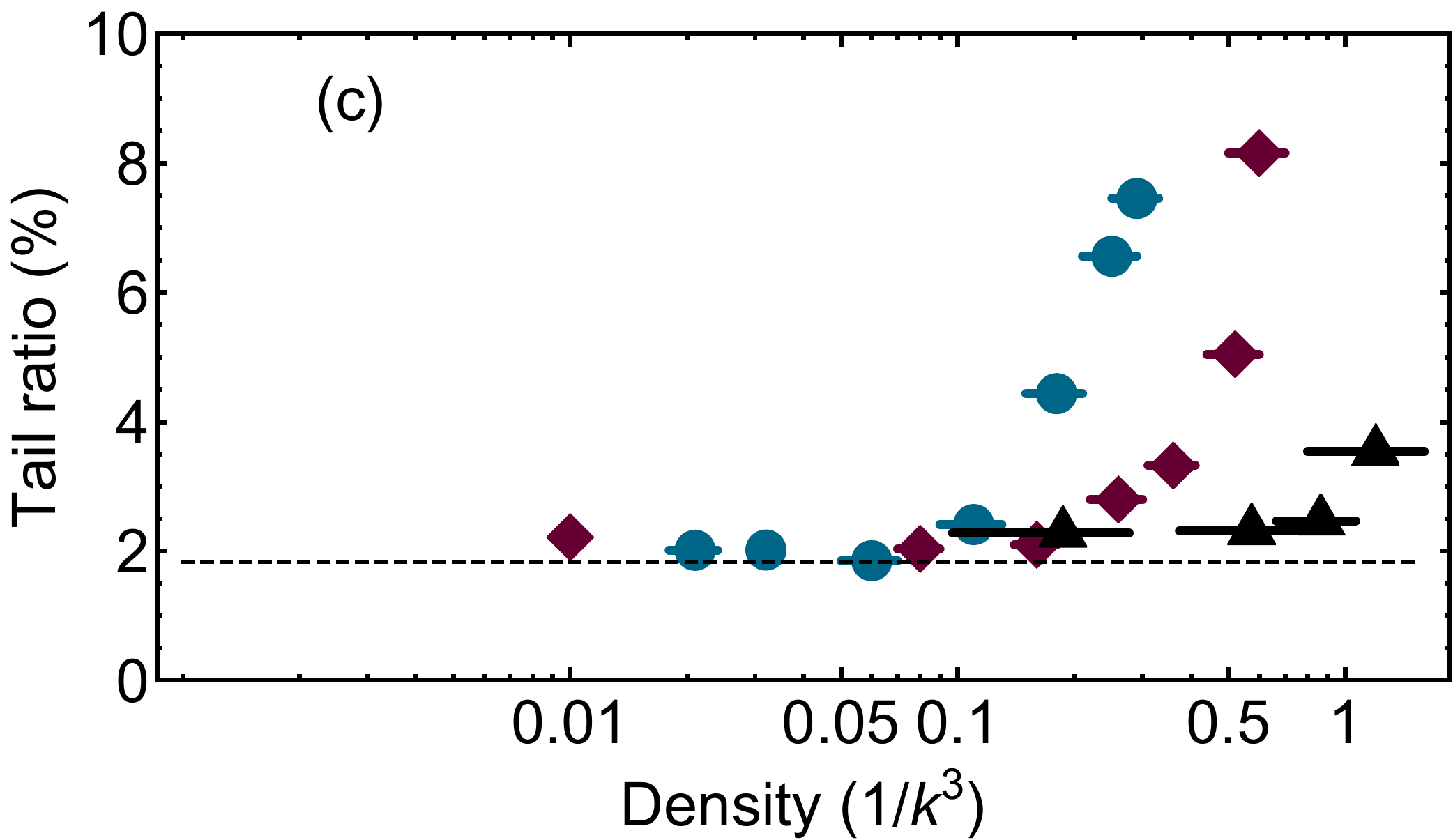} 
\caption{Tail ratio as defined by Eq.\,\eqref{eq3} for the same data sets as used to extract the decay times in Fig.\,\ref{fig2} ($s\simeq27$).}
\label{figtTRN}
\end{figure}

We analyze the behaviour of this tail ratio as a function of atom number on the same data as the one used for Fig.\,\ref{fig2}. In Fig.\,\ref{figtTRN} we indeed see the collapse as a function of atom number similarely to what found for the decay time (see discussion in Sec.\,\ref{sec:charac}).\par

%

\section{Measurements at large saturation intensity}\label{app:highI}
\begin{figure}[h!]
\includegraphics[width=0.85\linewidth]{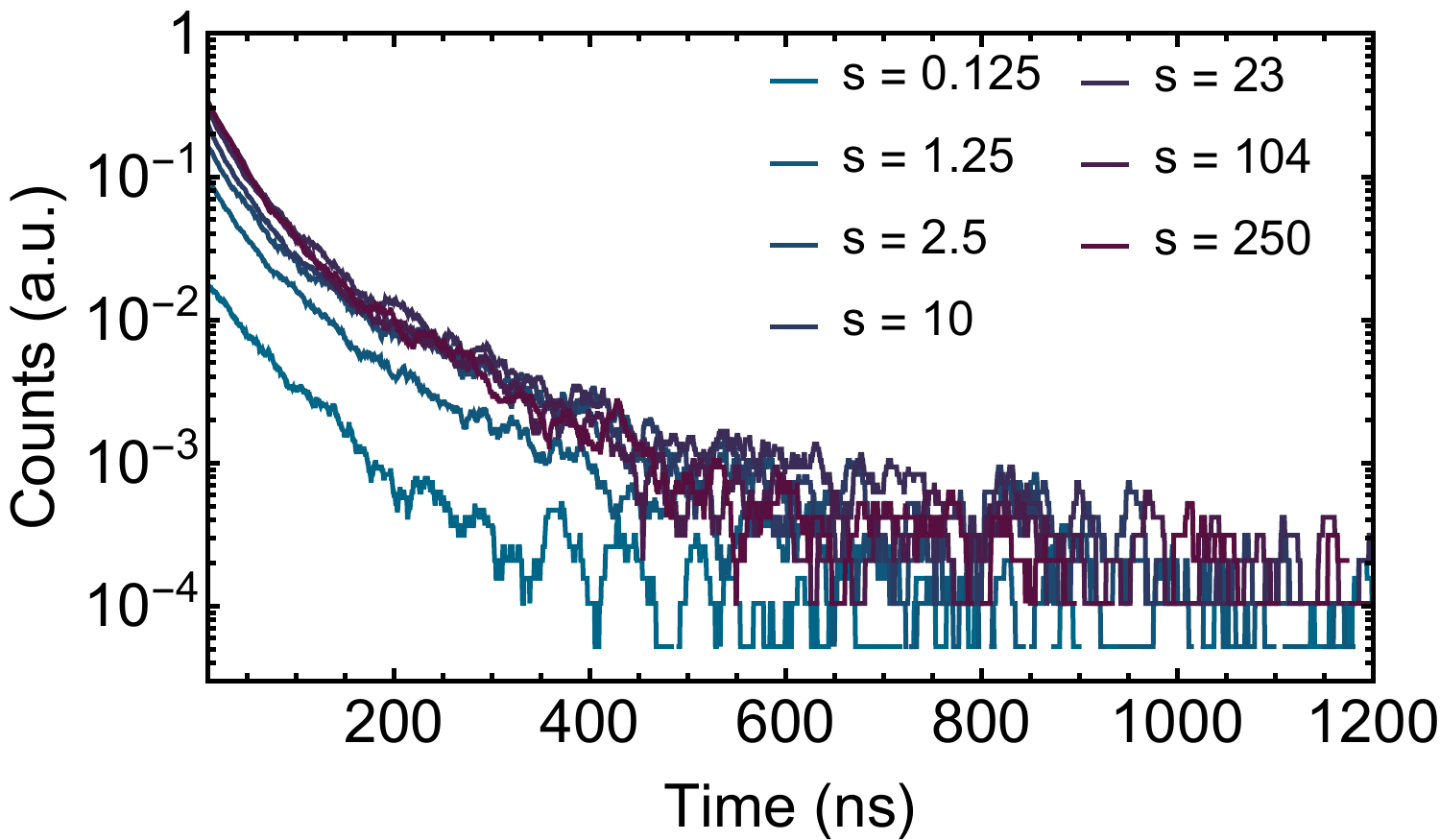} 
\includegraphics[width=0.85\linewidth]{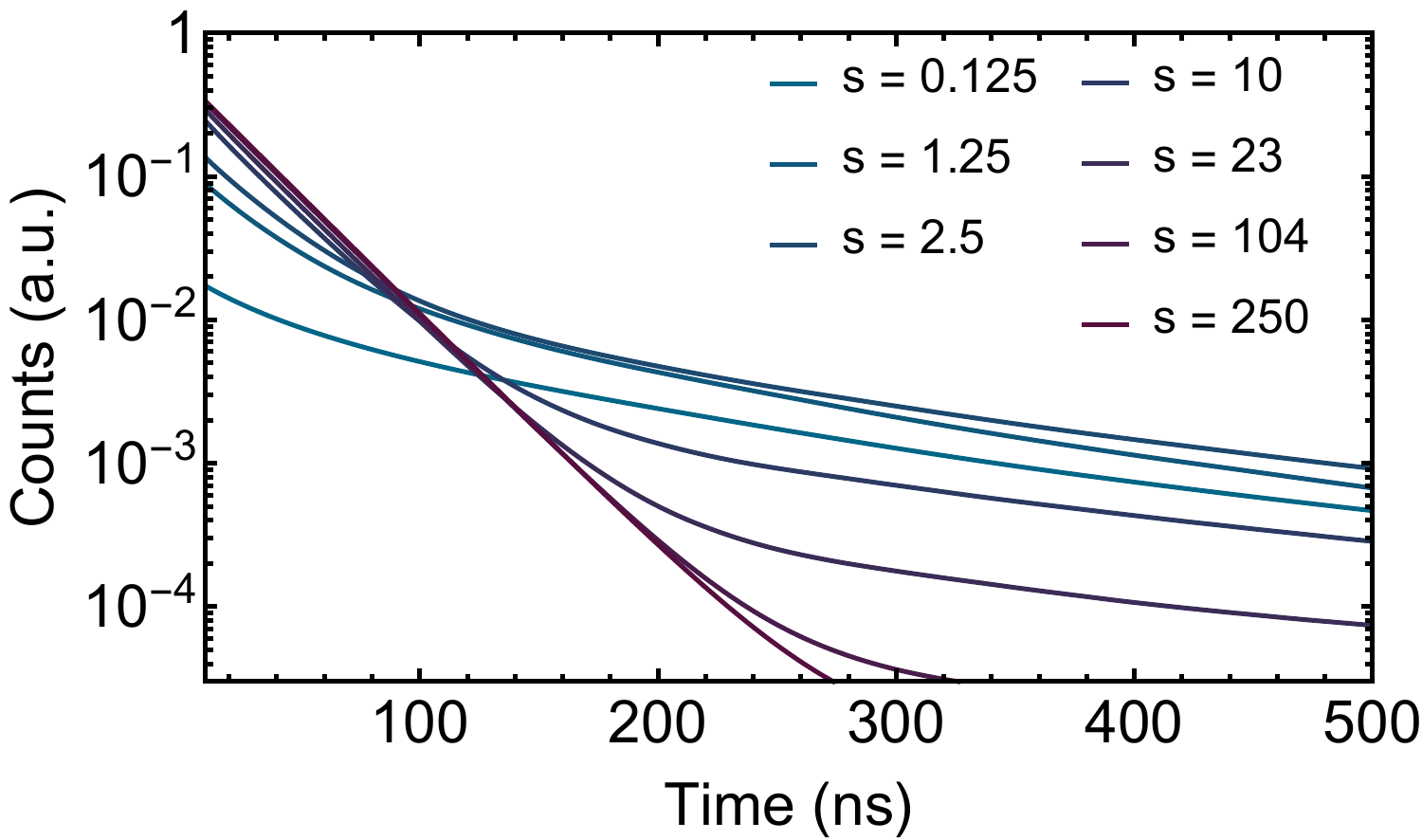} 
\caption{(a) Photon count decays acquired for different excitation intensities. (b) Numerical simulations performed with a non-linear coupled dipoles model, using 100 atoms at a density $\rho_0/k^3=0.3$. The traces have been normalized to the steady state value of the measurement at largest intensity. Here we define the origin of the time ($t=0$) $\SI{30}{\nano \second}$ after the switch off of the excitation. }
\label{figSI7}
\end{figure}

In this section we report the data acquired at very high values of the drive intensity, reaching $s=I/I_{\rm sat}\simeq 250$. The measurements have been performed with a larger atom number, $N\simeq \si{6000}$. For this reason this dataset and the one used for Fig.\,\ref{fig3} of main text cannot be directly compared. In  Fig.\,\ref{figSI7}(a) we report the temporal traces acquired in this way, normalized to the steady state value of the trace taken with the largest excitation intensity. In  Fig.\,\ref{figSI7}(b) we instead report the results of numerical simulations performed using the non-linear coupled dipoles model using 100 atoms distributed in a Gaussian cloud with a peak density $\rho_0=0.3/k^3$. The reported lines are the results of 10 realizations of the same numerical experiment. As one can see, according to the mean-field model, the subradiance is expected to disappear as the excitation strength is increased. The fact that experimentally the system still hosts a subradiant excitation even at very large intensity shows that the density matrix of the system cannot be factorized throughout the decay.\par
    
\section{Fluorescence decay curves}\label{app:curves}
\begin{figure*}[t]
\centering
    \includegraphics[width=0.5\linewidth]{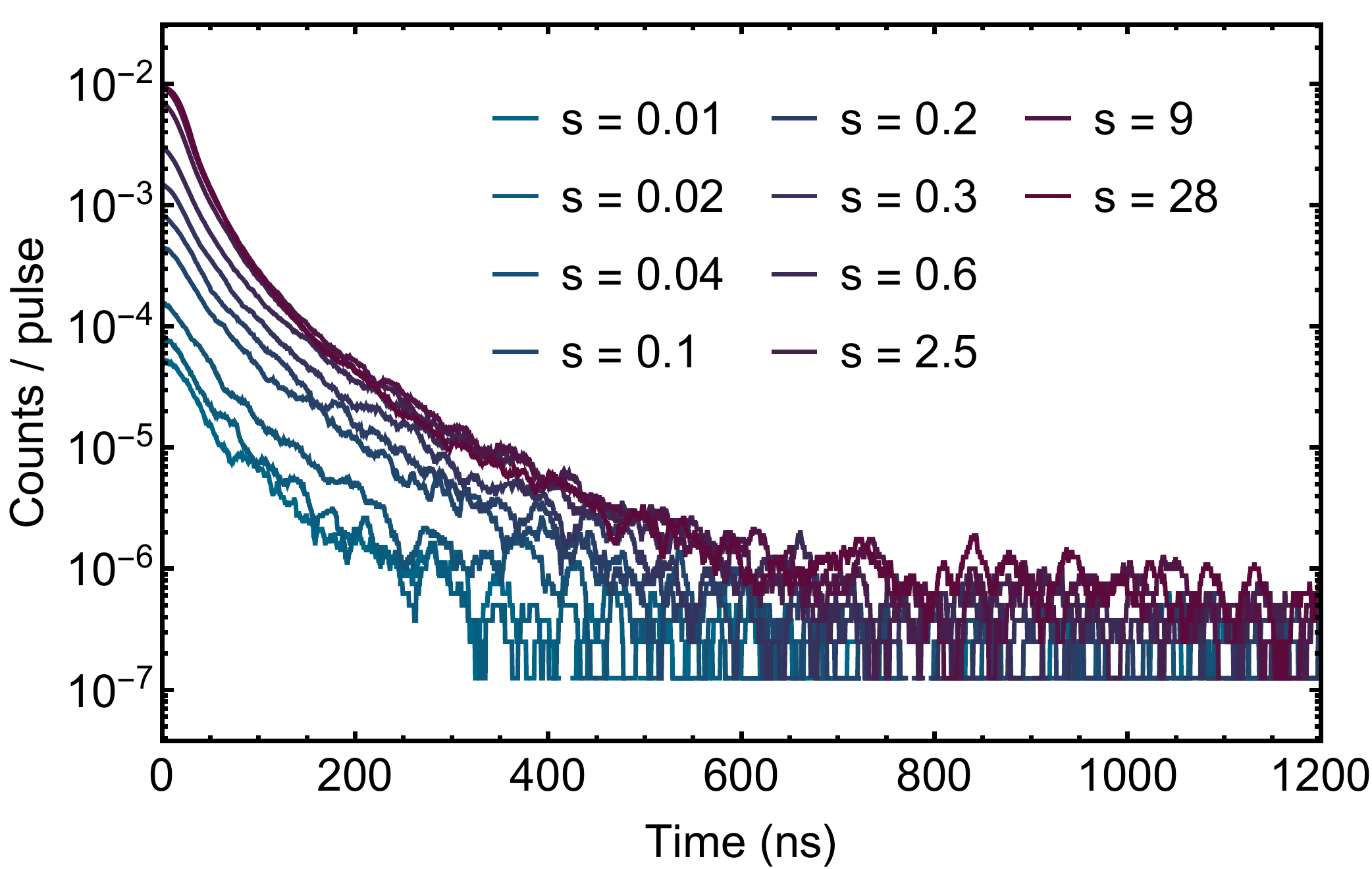}\hfil
    \includegraphics[width=0.5\linewidth]{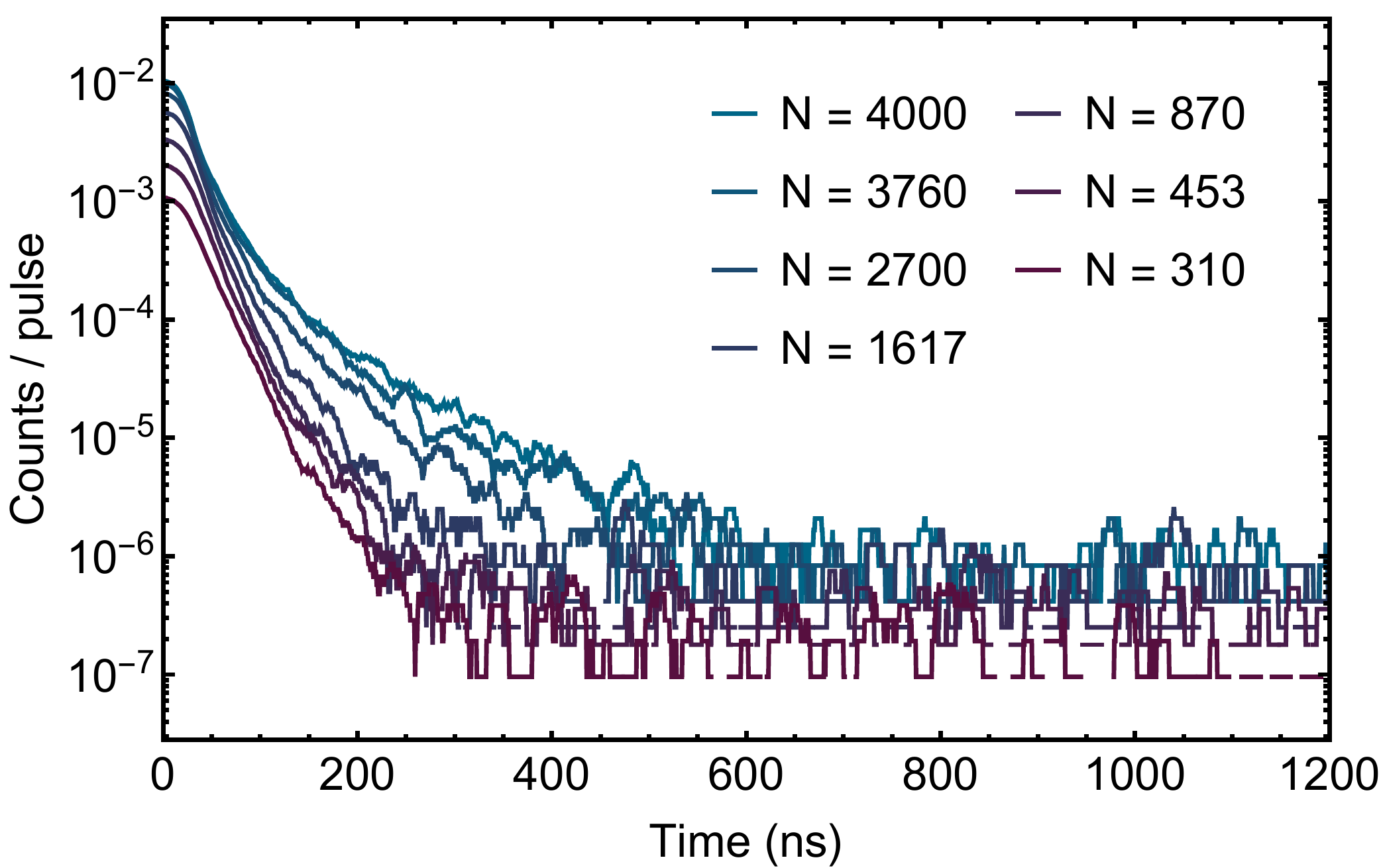}\par\medskip
\caption{Left panel: photon count decays acquired for different drive intensities. Right panel: photon count decays acquired as a function of the atom number for the first trapping geometry (circles in Figs.\,\ref{fig2} and \ref{figtTRN}).}
\label{figSI4}
\end{figure*}

For completeness, in  Fig.\,\ref{figSI4} we report the experimental photon count decay curves used in the analysis described in the main text. In the left panel the ones measured as a function of atom number and in the right panel, for different drive intensities.

\end{document}